\documentclass[reprint,amsmath,amssymb,aps,superscriptaddress,twocolumn,10pt]{revtex4-1}
\usepackage[colorlinks=true,allcolors=blue]{hyperref}
\usepackage{graphicx,color}
\usepackage{dcolumn}
\usepackage{subcaption}
\usepackage{bm}
\usepackage{amsmath,amsfonts,amssymb}
\usepackage{acronym}
\usepackage{enumitem}
\usepackage{array}
\usepackage{multirow}
\allowdisplaybreaks
\newcommand{\be}{\begin{equation}}
\newcommand{\ee}{\end{equation}}
\newcommand{\bea}{\begin{eqnarray}}
\newcommand{\eea}{\end{eqnarray}}

\newcommand{\rmd}{\mathrm{d}}

\newcommand{\BNU}{School of Physics and Astronomy, Beijing Normal University, Beijing 100875, China.}
\newacro{GW}{gravitational wave}
\newacro{MBHB}{massive black hole binary}
\newacro{BH}{black hole}
\newacro{SIS}{singular isothermal sphere}
\newacro{NFW}{Navarro-Frenk-White}
\newacro{PE}{parameter estimation}
\newacro{SNR}{signal-to-noise ratio}
\newacro{PN}{post newtonion}
\newacro{FIM}{Fisher Information Matrix}
\newacro{GWTC}{Gravitational-wave Transient Catalog}

\begin{document}

\title{Bayesian model selection of Primordial Black Holes and Dressed Primordial Black Holes with lensed Gravitational Waves}

\author{Xin-yi Lin}
\email{xinyilin@bnu.edu.cn}
\affiliation{\BNU}

\author{Zhengxiang Li}
\email{zxli918@bnu.edu.cn}
\affiliation{\BNU}

\author{Jian-dong Zhang}
\email{zhangjd9@mail.sysu.edu.cn}
\affiliation{MOE Key Labortory of TianQin Mission, TianQin Research Center for Gravitational Physics $\&$ School of Physics and Astronomy, Frontiers Science Center for TianQin, CNSA Research Center for Gravitational Waves, Sun Yat-sen University (Zhuhai Campus), Zhuhai 519082, China}

\date{\today}

\begin{abstract}

If particle dark matter (DM) and primordial black holes (PBHs) coexist, PBHs will be surrounded by particle DM, forming celestial objects known as dressed PBHs (dPBHs). These structures suggest a scenario in which PBHs and DM can exist simultaneously. However, in the high-frequency regime, the gravitational lensing effect of bare PBHs is similar to that of dPBHs. Ground-based gravitational wave (GW) detectors are particularly sensitive to high-frequency GW signals. In this regime, the lensing effect of a point-mass lens with a mass in the range of $10^{-1} \sim 10^2~M_{\odot}$ becomes significant. In this work, we incorporate dPBH models with GW observations and employ Bayesian inference techniques to distinguish PBHs from dPBHs. Using the third-generation ground-based GW detectors, Einstein Telescope (ET) and Cosmic Explorer (CE), as examples, we demonstrate that these detectors can effectively differentiate the lensing effects of dPBHs from those of PBHs across a broad frequency range. Furthermore, we find that with a larger black hole (BH) mass inside the surrounding particle DM, ET and CE can distinguish these two lensed models with even greater precision.

\end{abstract}

\maketitle

\section{Introduction}


On September 14, 2015, the ground-based gravitational wave detectors Advanced LIGO\cite{LIGOScientific:2014pky,LIGOScientific:2016emj} and Virgo\cite{VIRGO:2014yos} observed the first gravitational wave (GW) signal, GW150914\cite{Abbott:2016ob}, ever detected by humans. It is estimated that the advanced version of LIGO will detect between 3 and 100 inspiral events, with a detection range extending to hundreds of megaparsecs (Mpc)\cite{Narayan:1991fn,Blanchet:1995ez}. In addition to these two detectors, the baseline installation of KAGRA\cite{KAGRA:2020tym,Somiya:2011np,Aso:2013eba} was completed in 2019. The LIGO, Virgo, and KAGRA network, collectively referred to as the LVK Collaboration, has participated in the O1, O2, and O3 observing runs, detecting approximately $\mathcal{O}(100)$ gravitational wave signals\cite{KAGRA:2021vkt,LIGOScientific:2018mvr,LIGOScientific:2020ibl,LIGOScientific:2021aug,LIGOScientific:2021usb,Garron:2023gvd} from coalescing compact binaries to date.


Similar to electromagnetic signals, when a gravitational wave (GW) signal propagates through a celestial body, it experiences amplification, deflection, and time delay due to the gravitational field\cite{Ohanian:1974ys,Nakamura:1997sw,Nakamura:1999uwi,Schneider:2006,Takahashi:2003ix,Oguri:2019fix,Leung:2023lmq}. In the regime where the wavelength is much shorter than the Schwarzschild radius of the lens, known as the geometric-optics approximation, GW signals may be split into multiple distinct event pairs. This phenomenon is referred to as gravitational wave lensing. Extensive studies have been conducted on the gravitational lensing of GW signals\cite{Fan:2016swi,Broadhurst:2019ijv,Singer:2019vjs,McIsaac:2019use,Hannuksela:2019kle,Liu:2020par,Dai:2020tpj,LIGOScientific:2021izm,Diego:2021fyd,Baker:2016reh,Goyal:2020bkm,Lai:2018rto,Diego:2019rzc,Oguri:2020ldf,Xu:2021bfn,Tambalo:2022wlm,Caliskan:2022hbu,Urrutia:2021qak,Urrutia:2023mtk,Frampton:2010sw,Zhou:2022yeo,Huang:2023prq,Huang:2024zvk}. However, in the first three observing runs, no conclusive evidence has been found to confirm the existence of lensed GW signals. With the ongoing improvement in the sensitivity of GW detectors, the fourth observing run (O4), which consists of two phases, O4a and O4b, is scheduled to conclude its observational campaign on June 9, 2025. As of now, the number of detected GW events in O4 has reached 149\cite{Capote:2024rmo}, surpassing the total from O1 to O3. This suggests an increased likelihood of detecting lensed GW event pairs in the fourth observing run. Furthermore, third-generation observatories such as the Einstein Telescope (ET)\cite{Branchesi:2023mws,Punturo:2010zza} and Cosmic Explorer (CE)\cite{Reitze:2019iox} are expected to become operational in the near future. The advent of these next-generation GW detectors holds promising prospects for the observation of gravitational wave lensing effects.


Cold dark matter (CDM) is one of the most essential tools for studying the formation of structure in the Universe\cite{SpringelV:2005x}. In this work, we focus on primordial black holes (PBHs), which form from the collapse of large density fluctuations in the early Universe\cite{Hawking:1971ei,Meszaros:1975ef,Garcia-Bellido:1996mdl,Boybeyi:2024mhp} and are considered potential candidates for dark matter (DM)\cite{Hawking:1971ei,Carr:1974nx,Carr:1975qj,Chapline:1975ojl,Kawasaki:1997ju,Kohri:2007qn,Khlopov:2008qy,Frampton:2010sw,Carr:2021bzv,Green:2020jor,Wang:2021lij}. When particle DM coexists with PBHs, the PBHs will be surrounded by particle DM\cite{Bertschinger:1985pd,Mack:2006gz,Ricotti:2007au,GilChoi:2023ahp}, forming structures known as dressed PBHs (dPBHs). These PBHs grow by accreting DM halos from their surroundings until a redshift of $z\approx30$, at which large-scale structure formation occurs\cite{Mack:2006gz}. The presence of dPBHs implies the co-existence of particle DM and PBHs.

Ground-based gravitational wave (GW) detectors are particularly sensitive to GW signals at high frequencies. Within the frequency range of these detectors, the lensing effect of stellar-mass point-mass lenses becomes significant\cite{Nakamura:1997sw,Jung:2017flg,Lai:2018rto,Dai:2018enj,Oguri:2020ldf}. In the low-frequency diffraction regime, the amplification caused by dPBHs does not match that of PBHs. However, in the high-frequency geometric-optics regime, the amplification factors of PBHs and dPBHs can be nearly similar\cite{GilChoi:2023ahp}.

In this paper, we incorporate the PBH and dPBH lensing models within the frequency range of ground-based gravitational wave (GW) detectors and employ Bayesian statistics to distinguish these two models using third-generation detectors, Einstein Telescope (ET) and Cosmic Explorer (CE). We compute the Bayes factor and analyze the posterior distributions of the hyperparameters.


This paper is organized as follows. In Sec.\ref{Lensing effect of bare and dressed PBHs}, we examine the amplification factors of PBHs and dPBHs, comparing their respective effects. In Sec.\ref{Bayes factor}, we present our model selection methodology by introducing the Bayesian framework and analyzing the likelihood functions associated with different lensed models. In Sec.\ref{Result}, we present the posterior distributions of the hyperparameters for both lensed models. Additionally, we compute the Bayes factor for various black hole masses and source parameters. Finally, in Sec.\ref{Result}, we summarize our findings and discuss related implications. Throughout this work, we adopt the geometric unit system ($G = c = 1$).

\section{Lensing effect of bare and dressed PBHs}
\label{Lensing effect of bare and dressed PBHs}

At any given frequency, the lensed waveform $h_L(f)$ is related to the unlensed waveform through
\begin{equation}
h_L(f)=F(f)h_0(f),
\end{equation}
where $h_0(f)$ denotes the unlensed waveform. The amplification factor $F(f)$, defined as the ratio of the lensed to the unlensed waveform in the frequency domain, can be derived from the diffraction integral\cite{Takahashi:2003ix,Deguchi:1986zz,Nakamura:1999uwi,Leung:2023lmq}
\begin{equation}
F(f)=\frac{1}{2\pi i}\int \frac{\rmd^2\textbf{x}}{x^2_F},e^{i\,\phi(\textbf{x},\textbf{y})/{x^2_F}},
\end{equation}
where $\textbf{x}$ represents the physical coordinates that parameterize the two-dimensional lens plane, $\textbf{y}$ denotes the source position relative to the lens, and $x_F$ is the characteristic length scale of the lens system\cite{Oguri:2020ldf,Choi:2021bkx,Macquart:2004sh}
\begin{equation}
x_F=\sqrt{\frac{d_\mathrm{eff}}{2\pi f(1+z_L)}},
\end{equation}
where $d_\mathrm{eff} = d_L d_{LS} / d_S$. The propagation phase $\phi(\textbf{x},\textbf{y}) = \frac{1}{2} |\textbf{x} - \textbf{y}|^2 - \psi(\textbf{x}) - \phi_m(\textbf{y})$, also known as the time delay, is determined by the relative source position, the lensing potential $\psi(\textbf{x})$, and an additional phase term that ensures the minimum value of the time delay is zero.

The diffraction integral in $F(f)$ must be evaluated over the entire lensing plane. Moreover, the integrand is highly oscillatory, making it only conditionally convergent. Direct computation of the integrand is challenging, and achieving the desired precision would require substantial computational time. To enhance efficiency, we employ the asymptotic expansion method\cite{Guo:2020eqw}.

To compute $F(f)$ using the asymptotic expansion method, we introduce the dimensionless quantity
\begin{equation}
w=\frac{2\pi f(1+z_L)}{d_\text{eff}}\xi^2,
\end{equation}
where $\xi$ is the normalization constant for the length scale in the lens plane.

For any smoothly varying function $f(z)$, we have
\begin{equation}
\begin{aligned}
\label{asymptotic_expansion_method}
\int_{0}^{\infty}dz\,e^{iw z}f(z)&=\int_{0}^{b}dz\,e^{iw z}f(z)\\&+e^{iw b}\sum\limits_{n=1}\limits^{\infty}\frac{(-1)^n}{(iw)^n}\frac{\partial^{n-1}f}{\partial z^{n-1}}\Bigg|_{z=b}.
\end{aligned}
\end{equation}
And for axially symmetric lens, we have
\begin{equation}
F(w)= \int_{0}^{\infty}dz\,e^{iw z}f(z)
\end{equation}
where
\begin{equation}
\label{fz}
f(z)=-i w e^{i w y^2/2}J_0(w x_s \sqrt{2z})e^{-i w \psi(\sqrt{2z})}
\end{equation}
In this work, $n=5$ provides sufficient accuracy.

In the low-frequency regime, wave diffraction distorts both the amplitude and phase of the complex function $F(f)$. Therefore, we compute $F(f)$ using the asymptotic expansion method to handle the diffraction integral. However, in the intermediate and high-frequency regimes, we use the geometric optics approximation to accelerate the computation. In these regimes, the total amplification factor is the sum of the contributions from all geometric images\cite{Sun:2019ztn,Dai:2017huk,Cremonese:2021puh}, labeled as $j = 1, 2, \cdots$:
\begin{equation}
F_{\rm geo}(w) = \sum_j |\mu_j|^{1/2} e^{i\left(w \tau_j - \frac{\pi}{2} n_j \right)}.
\end{equation}
Here, the magnification factor at each image position $\textbf{x}_j$ is given by
\begin{equation}
\mu_j = \left[ \text{det}\left(\textbf{I} - \frac{\partial^2 \phi(\textbf{x}_j)}{\partial \textbf{x} \, \partial \textbf{x}}\right) \right]^{-1},
\end{equation}
where $\textbf{I}$ is a $2 \times 2$ identity matrix, $n_j$ is the Morse index, and $n_j = 0, 1, 2$ when the position of the $j$-th image $\textbf{x}_j$ is located at the minimum, saddle, and maximum points of $\phi(\textbf{x}, \textbf{x}_s)$, respectively\cite{Dai:2017huk,Cremonese:2021puh,Wang:2021kzt}.

However, in the intermediate to high-frequency regime, $F_{\rm geo}$ does not accurately approximate the amplification factor. To improve accuracy, we introduce the post-geometrical optics correction $\delta F$\cite{Tambalo:2022plm,Takahashi:2004mc}. This correction consists of two components: the correction to the geometric magnification of images, denoted as $\delta F_m$, and an additional contribution, $\delta F_c$, arising from the diffracted image caused by the cuspy lens center. The expression for $F_{\rm geo}$, incorporating the post-geometrical optics correction beyond the geometric optics limit, is given by
\begin{equation}
F(w)=\sum_j|\mu_j|^{1/2}\left(1+\frac{i}{w}\Delta_j\right) e^{i\,\left(w\,\tau_j-\frac{\pi}{2}\,n_j\right)}
\label{Fgeo1}
\end{equation}
where
\begin{equation}
\Delta_j=\frac{1}{16}\left[\frac{1}{2\alpha_j^2}\psi_j^{(4)}+\frac{5}{12\alpha_j^3}{\psi_j^{(3)}}^2+\frac{1}{\alpha_j^2}\frac{\psi_j^{(3)}}{|x_j|}+\frac{\alpha_j-\beta_j}{\alpha_j\beta_j}\frac{1}{|x_j|^2}\right]
\end{equation}
The second term in Eq.\ref{Fgeo1} is the correction to the magnification factor of geometric image
\begin{equation}
\delta F_m(w)=\frac{i}{w}\,\sum_j\,\Delta_j\,|\mu_j|^{1/2}\,e^{i\,(w\,\tau_j-\frac{\pi}{2}\,n_j)}
\end{equation}
The correction term $\delta F_c$ originates from the central density cusp of the lens. The form of $\delta F_c$ depends on the specific lens model.

\subsection{PBHs}
For a bare PBH, all of its mass is concentrated at a single point, allowing it to be modeled as a point-mass lens. The mass density is given by\cite{Tambalo:2022plm,Morita:2019sau}.
\begin{equation}
\rho(\textbf{r})=M_L\delta^3(\textbf{r}).
\end{equation}
And the surface mass density of PBH is $\Sigma(\textbf{x})=M_L\delta^2(\textbf{x})$ where $M_L$ is the PBH mass. The Einstein radius $x_E$ is $\sqrt{4M_L d_{LS}d_L/d_S}$. The dimensionless lensing potential is $\phi(\textbf{x})=ln|\textbf{x}|$. The amplification factor $F(f)$ for a bare PBH is\cite{Takahashi:2003ix}
\begin{equation}
\begin{split}
F(w)= & \exp\left[\frac{\pi w}{4}+\frac{i w}{2}\,\left(\ln \frac{w}{2}-2\phi_m(y) \right)\right]\\&\Gamma\left(1-\frac{i w}{2}\right)\,_1F_1\left(\frac{i w}{2},1,y^2\frac{i w}{2} \right),
\label{F_PBH}
\end{split}
\end{equation}
where $\phi_m(y)=(x_m-y)^2/2-\ln x_m$ with $x_m=(y+\sqrt{y^2+4})/2$. $\Gamma(z)$ is the Euler gamma function, and $_1F_1(a, b, z)$ is Kummer's confluent hypergeometric function.

In the geometric-optics regime, we consider the case where $w > 10$. The multiplicative factor is given by
\begin{equation}
\label{Fgeo_PBH}
F_{\rm geo}(w)=|\mu_+|^{1/2}-i\,|\mu_-|^{1/2}e^{i\,w\,\Delta \tau},
\end{equation}
where the magnification of the two geometric images are $\mu_{\pm}=1/2\pm(y^2+2)/(2\,y\,\sqrt{y^2+4})$, $\triangle \tau$ is the time delay between the two images and $\triangle\tau=y\sqrt{y^2+4}/2+\ln[(\sqrt{y^2+4}+y)/(\sqrt{y^2+4}-y)]$. For bare PBH, the term corresponding to the diffracted image $\delta F_c$ is zero\cite{Takahashi:2004mc}. Thus $\delta F$ is only contributed by the post-geometric correction to the amplification of the geometric images $\delta F_m$. Here
\begin{equation}
\begin{split}
\delta F(w)&=\frac{i}{3\,w}\frac{4x_+^2-1}{(x_+^2+1)^3(x_+^2-1)}|\mu_+|^{1/2}\\&+\frac{1}{3\,w}\frac{4x_-^2-1}{(x_-^2+1)^3(x_-^2-1)}|\mu_-|^{1/2}\,e^{i w\Delta T},
\end{split}
\end{equation}
where $x_{\pm}=(y\pm\sqrt{y^2+2})/2$ are the positions of the two geometric images.

\subsection{dPBHs}
During the radiation-dominated era, the halo mass increases to the point where the ratio $m_h / m_\text{PBH} \simeq 1$. In contrast, during the matter-dominated era, the halo mass grows in accordance with the cosmological expansion, $\propto (1 + z)^{-1}$. This growth stops at the redshift $z_c \sim 30$, where large-scale structure formation occurs. As a result, the halo mass is\cite{Mack:2006gz,Adamek:2019gns,Berezinsky:2013fxa,Boudaud:2021irr}
\begin{equation}
m_h=97\left(\frac{31}{1+z_c}\right)m_\text{PBH}
\end{equation}
enclosed within $R_h=0.61\,pc(31/(1+z_c))(m_h/M_{\odot})^{1/3}$\cite{Bertschinger:1985pd,Berezinsky:2013fxa,Boudaud:2021irr}. The density of halo is\cite{Serpico:2020ehh,Lacki:2010zf,Boudaud:2021irr}
\begin{equation}
\rho_h(r)=0.26M_{\odot}pc^{-3}((1+z_c)/31)^3(R_h/r)^{9/4}
\end{equation}
And the surface mass density of dPBH is\cite{Oguri:2022fir,GilChoi:2023ahp}
\begin{equation}
\begin{split}
\Sigma_h(\textbf{x})&=2\int_0^\infty dz\rho_h(\sqrt{\textbf{x}^2+z^2})\\&\simeq\rho_0 R_h\sqrt{\pi}\frac{\Gamma(5/8)}{\Gamma(9/8)}\left(\frac{R_h}{|\textbf{x}|}\right)^{5/4}
\end{split}
\end{equation}

The lensing potential can be divided into two parts. Then we have $\psi(\textbf{x})=\psi_h(\textbf{x})+\psi_\text{PBH}(\textbf{x})$\cite{GilChoi:2023ahp}, where
\begin{equation}
\psi_h(\textbf{x})=\frac{32}{9}\frac{\rho_0R_h^3}{\Sigma_{crit}}\pi^{1/2}\frac{\Gamma(5/8)}{\Gamma(9/8)}\left(\frac{|\textbf{x}|}{R_h}\right)^{3/4}
\end{equation}

For a dressed PBH, although the halo mass is much larger than that of a PBH surrounded by the halo, the mass enclosed within the Einstein radius makes the major contribution to lensing. The Einstein radius of a dPBH is approximately\cite{GilChoi:2023ahp}
\begin{equation}
x_E \simeq x_E^\text{PBH}\left[1+4.1\left(\frac{m_\text{PBH}}{M_{\odot}}\right)^{1/5}\left(\frac{d_{eff}}{Gpc}\right)^{3/5}\right]^{1/2}
\end{equation}
where $x_E^\text{PBH}$ is the Einstein radius of a bare PBH.

Since there is no analytical result for the amplification factor for dPBHs, in the wave diffraction regime where $w < 6$, we calculate the diffraction integral numerically using the asymptotic expansion method introduced earlier. As $\psi(\textbf{x})$ for dPBHs is not dimensionless, we define a new quantity, $w_0 = x_E^2 w$, to calculate the amplification factor using the asymptotic expansion method. Then, the function $f(z)$ in Eq.\ref{fz} can be written as
\begin{equation}
f(z)=-i\frac{w_0}{x_E^2}e^{i w_0 (\frac{y}{x_E})^2/2}J_0(w_0 \frac{y}{x_E} \frac{\sqrt{2z}}{x_E})e^{-i w_0 \frac{\psi(\sqrt(2z)}{x_E^2}}.
\end{equation}
Now we make $z'=z/x_E^2$, we have
\begin{equation}
\begin{split}
f(z')&=-i\frac{w_0}{x_E^2}e^{i w_0 y^2/2} J_0(w_0 y \sqrt{2z'})e^{-i w_0 \frac{\psi(\sqrt{2z})}{x_E^2}}\\
&=-i\frac{w_0}{x_E^2}e^{i w_0 y^2/2} J_0(w_0 y \sqrt{2z'})\\
&\times e^{-i w_0 [c_1\frac{(x_E\sqrt{2z'})^{3/4}}{x_E^2}+c_2\frac{\ln(x_E\sqrt{2z'})}{x_E^2}]},
\end{split}
\end{equation}
where $c_1=\frac{32}{9}\frac{\rho_0 R_h^3}{\Sigma_{crit}}\sqrt{\pi}\frac{\Gamma(5/8)}{\Gamma(9/8)}$, $c_2=\frac{m_\text{PBH}}{\pi \Sigma_{crit}}$. Then we have
\begin{equation}
\begin{split}
dz\cdot f(z)\cdot e^{iwz}&=-iw_0\text{dz'}e^{iw_0y^2/2} J_0(w_0 y \sqrt{2z'})e^{i w_0 z'}\\
&\times e^{-i w_0 [c_1\frac{(x_E\sqrt{2z'})^{3/4}}{x_E^2}+c_2\frac{\ln(x_E\sqrt{2z'})}{x_E^2}]}.
\end{split}
\end{equation}
Then we can define a new function
\begin{equation}
\begin{split}
f_1(z')&=-i w_0 e^{i w_0 y^2/2} J_0(w_0 y \sqrt{2z'})\\
&\times e^{-i w_0 [c_1\frac{(x_E\sqrt{2z'})^{3/4}}{x_E^2}+c_2\frac{\ln(x_E\sqrt{2z'})}{x_E^2}]},
\end{split}
\end{equation}
and we have
\begin{equation}
\begin{split}
F(w)&=\int_0^\infty dz\, f(z)\,e^{iwz}\\
&=\int_0^\infty dz'\,f_1(z')\,e^{iw_0z'}.
\end{split}
\end{equation}
Then we can calculate $F(w)$ by the asymptotic expansion in Eq.\ref{asymptotic_expansion_method}.

Thus, two images are formed on the lensing plane, and the amplification factor in the geometric-optics limit takes the same form as in Eq.\ref{Fgeo_PBH}. Due to the complexity of the dPBH lens model, we do not consider the post-geometrical optics correction here. When $w_0 > 6$\textbf{,} the accuracy of $F_{\rm geo}$ is sufficient.

\begin{figure}[h]
    \centering
    \begin{subfigure}[b]{0.5\textwidth}
        \includegraphics[width=\linewidth]{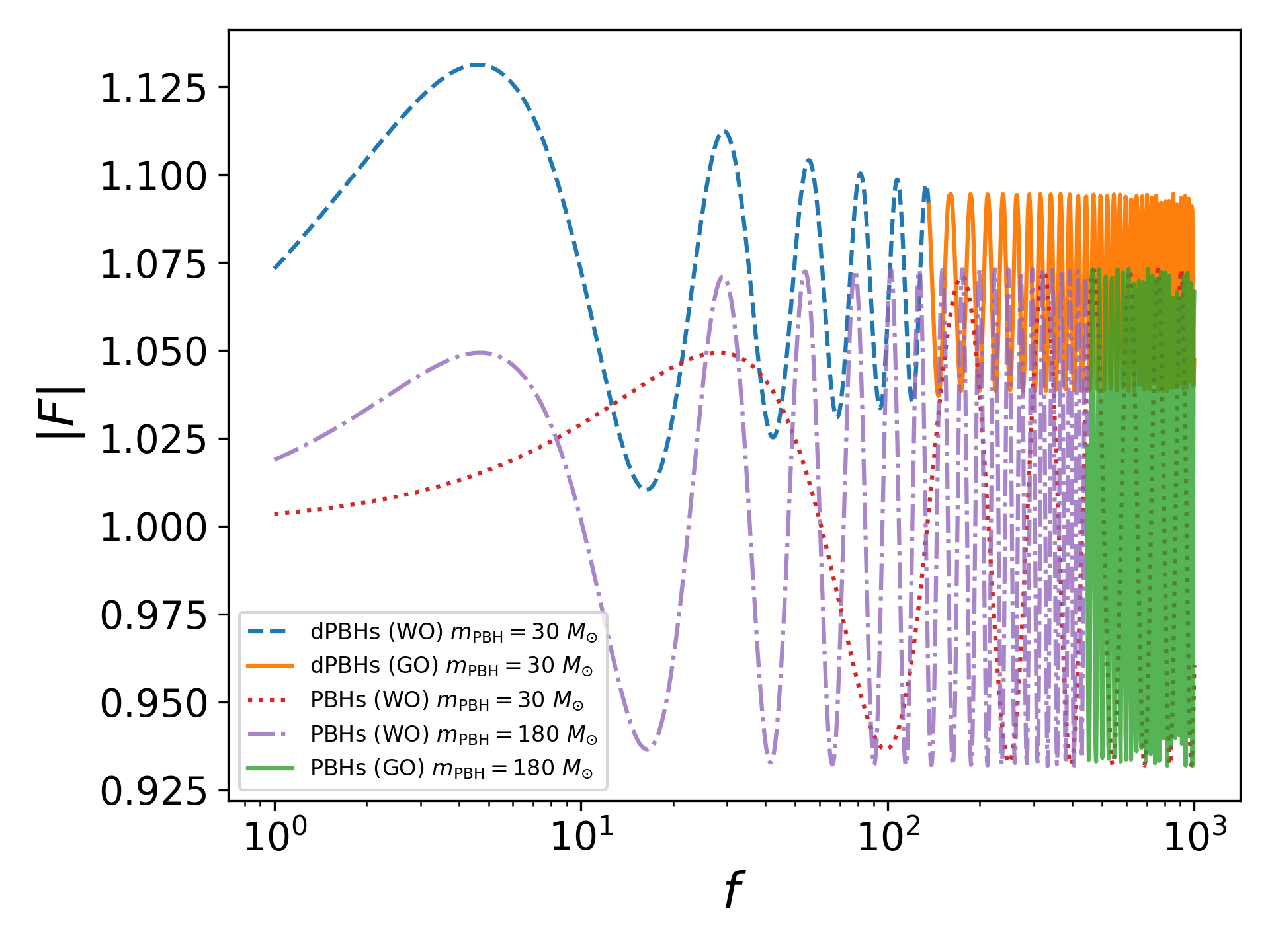}
        \caption{}
        \label{fig:Fw_abs}
    \end{subfigure}
    \hfill
    \begin{subfigure}[b]{0.5\textwidth}
        \includegraphics[width=\linewidth]{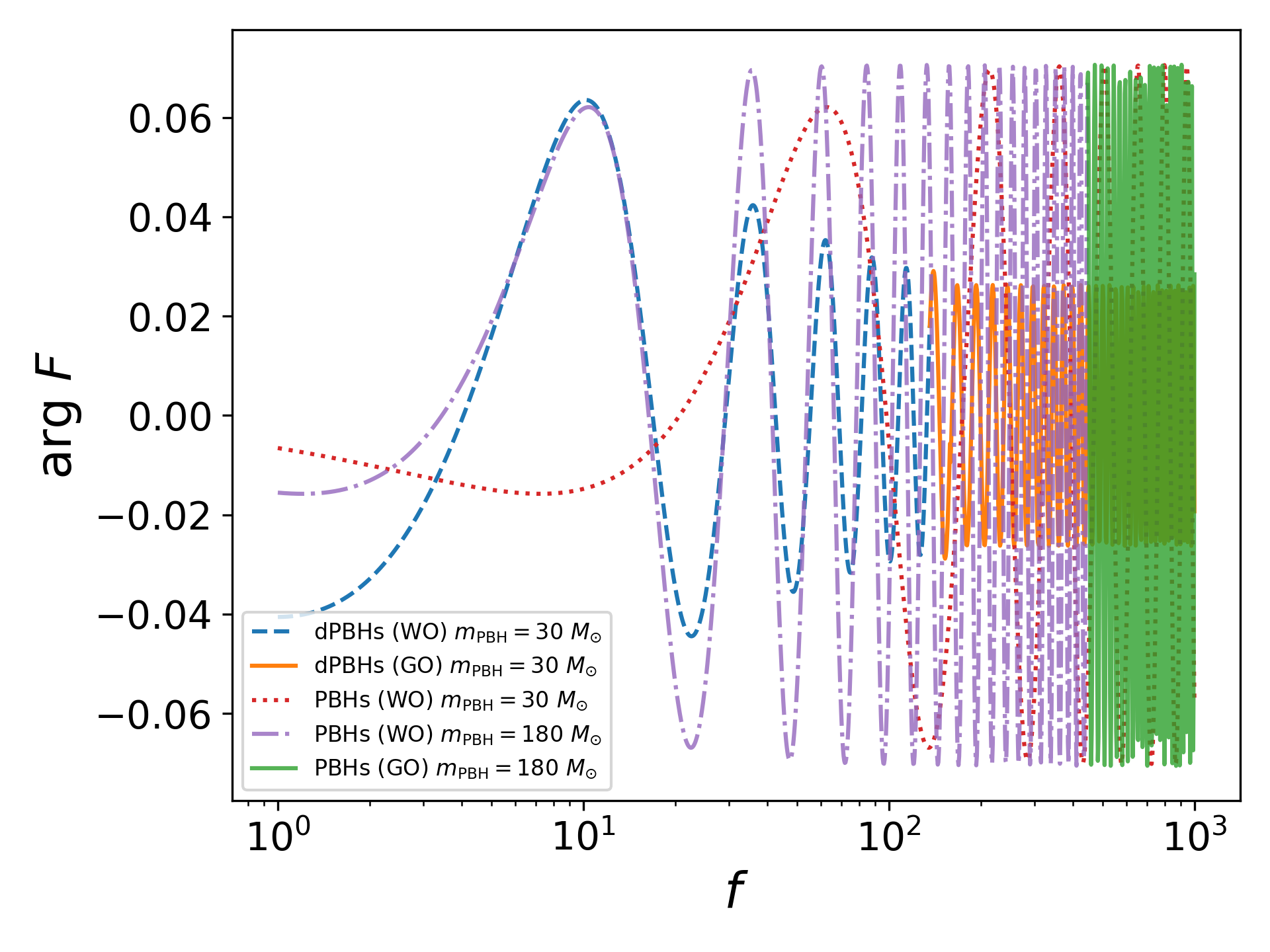}
        \caption{}
        \label{fig:Fw_arg}
    \end{subfigure}
    \caption{The amplification factors $F$ as a function of frequency $f$ for both dPBHs and PBHs across different regimes. The upper panel illustrates the absolute value $|F(f)|$, while the lower panel displays the corresponding phase $\text{arg}~F$. The dashed blue curves show $F(f)$ for dPBHs with $M_{\mathrm{PBH}} = 30~M_{\odot}$ in the wave-optics regime ($w_0 \leq 6$), while the solid orange curves reflect the geometric-optics regime ($w_0 > 6$). The dotted red curves depict PBHs with $M_{\mathrm{PBH}} = 30~M_{\odot}$ in the wave-optics regime. The dashdot purple curves display $F(f)$ for PBHs with $M_{\mathrm{PBH}} = 180~M_{\odot}$ in wave-optics ($w \leq 10$), in contrast to the solid green curves showing the geometric-optics behavior ($w > 10$).}
    \label{fig:Fw}
\end{figure}

In Fig. \ref{fig:Fw}, we present the amplification factors $F(f)$ for both dPBHs and PBHs across different regimes. We compute the amplification factors for these two models using the methods described above.  We consider two cases that $M_{\mathrm{PBH}} = 30~M_{\odot}$ for dPBHs and PBHs with $z_S = 1.5$, $z_L = 0.2$, and $y = 3.5$ and $M_{\mathrm{PBH}} = 180~M_{\odot}$ for PBHs with the same $z_S$, $z_L$, and $y$ values. For our subsequent analysis focusing on dimensional frequency dependence, we use $f$ rather than the dimensionless frequency $w$ for the $x$-axis.

As shown in Fig. \ref{fig:Fw}, it is clear that the amplification factors for dPBHs and PBHs with $M_{\mathrm{PBH}} = 30~M_{\odot}$ are significantly different. Therefore, we need to consider the alternative case where the amplification factors for dPBHs resemble those of PBHs. Notably, the amplitude $F(f)$ for dPBHs with $M_{\mathrm{PBH}} = 30~M_{\odot}$ closely matches that of PBHs with $M_{\mathrm{PBH}} = 180~M_{\odot}$, providing a valuable reference range for subsequent prior selection.

\section{Model selection}
\label{Bayes factor}

We assume that the ground-based detectors CE and ET observe the two images caused by lensing. We model the lens as a dPBH. Using Bayes' theorem, we compute the posterior distribution $p(\theta | d, H)$ for the source and lensed parameters $\theta$, given the data $d$ under model $H$.
\begin{equation}
p(\theta|d,H)=\frac{p(\theta|H)p(d|\theta,H)}{p(d|H)}.
\end{equation}
Under the Bayesian inference framework, the posterior is the product of the prior $p(\theta | H)$ and the likelihood $p(d | \theta, H)$, normalized by the evidence $p(d | H)$. The likelihood function, denoted as $p(\theta | H) = \mathcal{L}$, describes the probability of the observed data given the model parameters. The likelihood of the GW signal can be written in the form of an inner product

\begin{equation}
p(d|\theta,H)\propto\text{exp}\left[-\frac{1}{2}\left\langle d-h(\theta)|d-h(\theta)\right\rangle\right]
\end{equation}
where the inner product between two strain time series $a(t)$ and $b(t)$ is defined as\cite{Finn:1992wt,Cutler:1994ys}
\begin{equation}
(a|b)=4\,\mathfrak{Re}\,\int_0^{\infty}{\rm d}f\,\frac{\tilde{a}(f)\,\tilde{b}^*(f)}{S_N(f)},
\end{equation}
where $\tilde{a}(f)$ and $\tilde{b}(f)$ are the Fourier transforms of $a(t)$ and $b(t)$, respectively. The symbol ${}^*$ denotes the complex conjugate, and $S_N(f)$ represents the one-sided power spectral density (PSD). In this work, we use the PSD formulations for ET\cite{ET2023,Branchesi:2023mws,Punturo:2010zza} and CE\cite{CEsensitivity2023,Reitze:2019iox}.

We compute the evidence ratio to study which lens model is more likely to preferred by the observed data,
\begin{equation}
\text{BF}=\frac{\mathcal{Z}_{\text{dPBH}}}{\mathcal{Z}_{\text{PBH}}}
\end{equation}
where BF is the Bayes factor, and $\mathcal{Z}_{\text{dPBH}}$ and $\mathcal{Z}_{\text{PBH}}$ are the evidence for the dressed and bare PBH, respectively. $\mathcal{Z}_{\text{dPBH}}$ and $\mathcal{Z}_{\text{PBH}}$ can be written as
\begin{equation}
\begin{split}
\mathcal{Z}_{\text{dPBH}}&=\int\mathcal{L}(d|\theta,H_{\text{dPBH}})\\
\mathcal{Z}_{\text{PBH}}&=\int\mathcal{L}(d|\theta,H_{\text{PBH}})
\end{split}
\end{equation}
where $H_\text{dPBH}$ is the dPBH lensing detection hypothesis, $H_\text{PBH}$ is the PBH lensing detection hypothesis.

If the Bayes factor is positive, the data favors $H_{\text{dPBH}}$ over $H_{\text{PBH}}$. Conversely, if the Bayes factor is negative, the data prefers $H_{\text{PBH}}$. However, in practice, we typically use $\log$ for model selection,
\begin{equation}
\text{log}\text{BF}_\text{{PBH}}^{\text{dPBH}}=\text{log}\mathcal{Z}_{\text{dPBH}}-\text{log}\mathcal{Z}_\text{{PBH}}
\end{equation}
In this work, we adopt a threshold of $\log \text{BF} = 8$\cite{Thrane:2018qnx,Sun:2024nut}, meaning that if $\log \text{BF}$ reaches 8, the data is considered to support the dPBH lens model.

We use \texttt{dynesty}\cite{Speagle:2019ivv,zenodo2023}, which is based on the nested sampling method\cite{Skilling:2006gxv}, to perform Bayesian inference. With \texttt{dynesty}, we can simultaneously obtain both the evidence and the posterior probability distributions of parameters. Additionally, we use the phenomenological waveform model IMRPhenomD\cite{Khan:2015jqa} to characterize the waveform, incorporating the inspiral, merger, and ringdown.


\section{Result}
\label{Result}
In this section, we present the posterior distribution of the source and lens parameters. Subsequently, we discuss the validity of Bayesian model selection in distinguishing dPBHs from PBHs. Finally, to simulate a more realistic scenario, we consider variations in the source parameters. We then examine the dependence of the Bayes factor on the source parameters $\eta$ and $M$.

We assume that the GW signal is affected by the dPBH lens. Since we do not have additional prior information about these parameters, we adopt flat prior distributions. The priors for the intrinsic parameters, i.e., $\{\eta, M, t, z_S\}$, are set around their injected values. For the extrinsic parameters, i.e., $\{\phi_c, \theta_s, \phi_s, \theta_l, \phi_l\}$, we sample over the entire parameter space. Given that our primary focus is on the lens mass, we assume some prior knowledge of the lens properties. Consequently, we fix the lens redshift at $z_L = 0.2$ and the dimensionless source position at $y = 3.5$.

\subsection{The posterior distribution with dPBH and PBH}

We inject the data with a symmetric mass ratio of $\eta = 0.23$, a total mass of $M = 50~M_{\odot}$, an observation duration of $t = 1000~\text{s}$, a phase of coalescence of $\phi_c = 0$, and a redshift of the MBHB of $z_S = 1.5$. Additionally, we set all angle parameters to $\pi/3$ and assume the black hole halo mass embedded in the primordial halos of $M_{\text{PBH}} = 30~M_{\odot}$. The injected values, priors, and parameter estimates for the source and lens parameters are summarized in Table \ref{injected_values_and_priors}. The last two columns of Table \ref{injected_values_and_priors} show the parameter estimation precision at the $1\sigma$ confidence level for $H_{\text{dPBH}}$ and $H_{\text{PBH}}$.

It is worth noting that we adopt different priors for $M_{\text{PBH}}$ under different lensing detection hypotheses. The frequency range of ground-based detectors is sensitive to PBH masses in the range of $10^{-1} \sim 10^2 ~M_{\odot}$. To ensure an apparent lensing effect and just to be sure, we set the prior of $M_{\text{PBH}}$ for $H_{\text{PBH}}$ within $10 \sim 300~ M_{\odot}$. We have verified that the amplification of a dPBH with $M_{\text{PBH}} = 30~M_{\odot}$ is similar to that of a PBH with $M_{\text{PBH}} = 180~M_{\odot}$, indicating that the same prior is not required for these two lensing detection hypotheses. To compute the posterior distribution and evidence more efficiently, the prior range for $H_{\text{dPBH}}$ does not need to be as wide as that for $H_{\text{PBH}}$. Thus, we set the prior range of $H_{\text{dPBH}}$ to $10 \sim 50~M_{\odot}$.

\begin{table*}[ht]
\centering
\begin{tabular}{c c c c c c}
\hline
\textbf{Parameter} & \textbf{Symbol} & \textbf{Injected value}& \textbf{Prior} & \textbf{dPBH} & \textbf{PBH} \\
\hline
Symmetric mass ratio& $\eta$& 0.23& [0.15, 0.25]& $0.2293^{+0.0040}_{-0.0037}$& $0.2498^{+0.0001}_{-0.0003}$ \\
\hline
Total mass ($\text{M}_{\odot}$)& $M$& 50& [10, $10^3$]& $48.1442^{+3.5159}_{-3.4483}$& $24.7552^{+0.6520}_{-0.5963}$ \\
\hline
Observation duration (s)& t& 30& [10, 45]& $37.1626^{+5.2937}_{-9.0854}$& $11.9859^{+0.0363}_{-0.0157}$\\
\hline
The phase of coalescence& $\phi_c$& 0& [0, $2\pi$]& $5.1813^{+0.4938}_{-1.4534}$& $2.8203^{+0.0890}_{-0.0643}$\\
\hline
Source redshift& $z_S$& 1.5& [0.5, 5]& $1.6045^{+0.1950}_{-0.1736}$& $3.8207^{+0.1170}_{-0.1221}$\\
\hline
Source altitude& $\theta_s$& $\pi/3$& [0, $\pi$]& $1.3187^{+0.2224}_{-0.6120}$& $0.1400^{+0.1480}_{-0.0974}$\\
\hline
Source azimuthal& $\phi_s$& $\pi/3$& [0, $2\pi$]& $5.2448^{+0.3653}_{-0.5344}$&$0.3896^{+0.4653}_{-0.2742}$\\
\hline
Source angular momentum altitude& $\theta_l$& $\pi/3$& [0, $\pi$]& $2.4341^{+0.4296}_{-0.6399}$&$2.2913^{+0.1104}_{-0.0952}$\\
\hline
Source angular momentum azimuthal& $\phi_l$& $\pi/3$& [0, $2\pi$]& $0.5835^{+3.6165}_{-0.4069}$&$3.7802^{+0.8677}_{-0.8636}$\\
\hline
\multirow{2}{*}{BH mass ($\text{M}_{\odot}$)} & \multirow{2}{*}{$m_{\text{dPBH}}$}& \multirow{2}{*}{30}& [10, 50] for $H_{\text{dPBH}}$& \multirow{2}{*}{$29.3388^{+1.3090}_{-1.4634}$}& \multirow{2}{*}{$299.0500^{+0.7153}_{-1.3856}$}\\
& & & [10, 300] for $H_{\text{PBH}}$& & \\
\hline
\end{tabular}
\caption{The injected values, the priors and parameter estimation for the source and lens parameters. The last two columns show the parameter estimation precision at the $1\sigma$ confidence level for $H_{\text{dPBH}}$ and $H_{\text{PBH}}$.}
\label{injected_values_and_priors}
\end{table*}

\begin{figure}[h]
\centering
\includegraphics[width=0.5\textwidth]{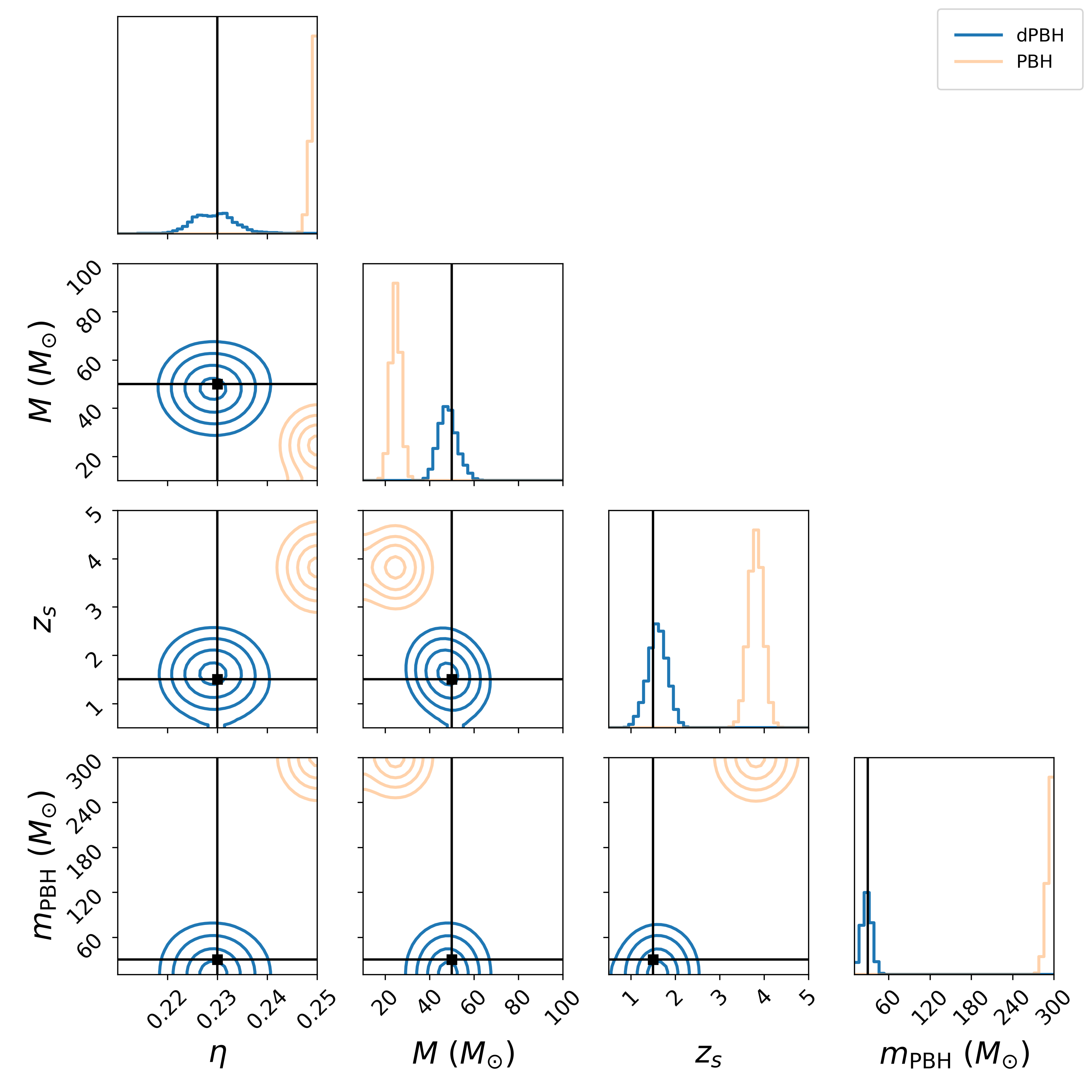}
\caption{The posterior distribution of $\eta$, $M$, $z_S$, and $M_{\text{PBH}}$. The blue curves and the ligth orange curves correspond to the dPBH detection hypothesis and the PBH detection hypothesis, respectively. The black lines indicate the injected values.} 
\label{result_dPBH_eta23_m50_ml30}
\end{figure}

To better analyze the results, we retain only four parameters, $\eta$, $M$, $z_S$, and $M_{\text{PBH}}$, in the corner plot. Fig. \ref{result_dPBH_eta23_m50_ml30} presents the posterior distribution of the key parameters. Comparing the contour curves in Fig. \ref{result_dPBH_eta23_m50_ml30}, we observe that the parameter estimation results for $H_{\text{dPBH}}$ are very close to the injected values, whereas those for $H_{\text{PBH}}$ exhibit noticeable deviations. Referring to the parameters in Table \ref{injected_values_and_priors}, the estimated values of $\eta$ and $m_{\text{dPBH}}$ for $H_{\text{dPBH}}$ closely match the injected values, while the estimations of $M$ and $z_S$ for $H_{\text{dPBH}}$ are fairly accurate.

The error estimation for the observation duration is not highly precise, as the GW signal is too weak at the time that far from the coalescence phase, thus contributing minimally. Similarly, the error estimations for the angular parameters are less reliable; however, these parameters are not the most critical factors in GW signal analysis. 

On the other hand, the parameter estimations for all parameters under $H_{\text{PBH}}$ exhibit significant deviations. The logarithm of the Bayes factor is found to be $\log \text{BF}^{\text{dPBH}}_{\text{PBH}} = 74.4319$, which is significantly greater than the threshold of $8$. This result strongly supports the validity of Bayesian analysis for lensed signals.

\subsection{The variety of Bayes factors with the source parameters}

\begin{figure}[h]
\centering
\includegraphics[width=0.5\textwidth]{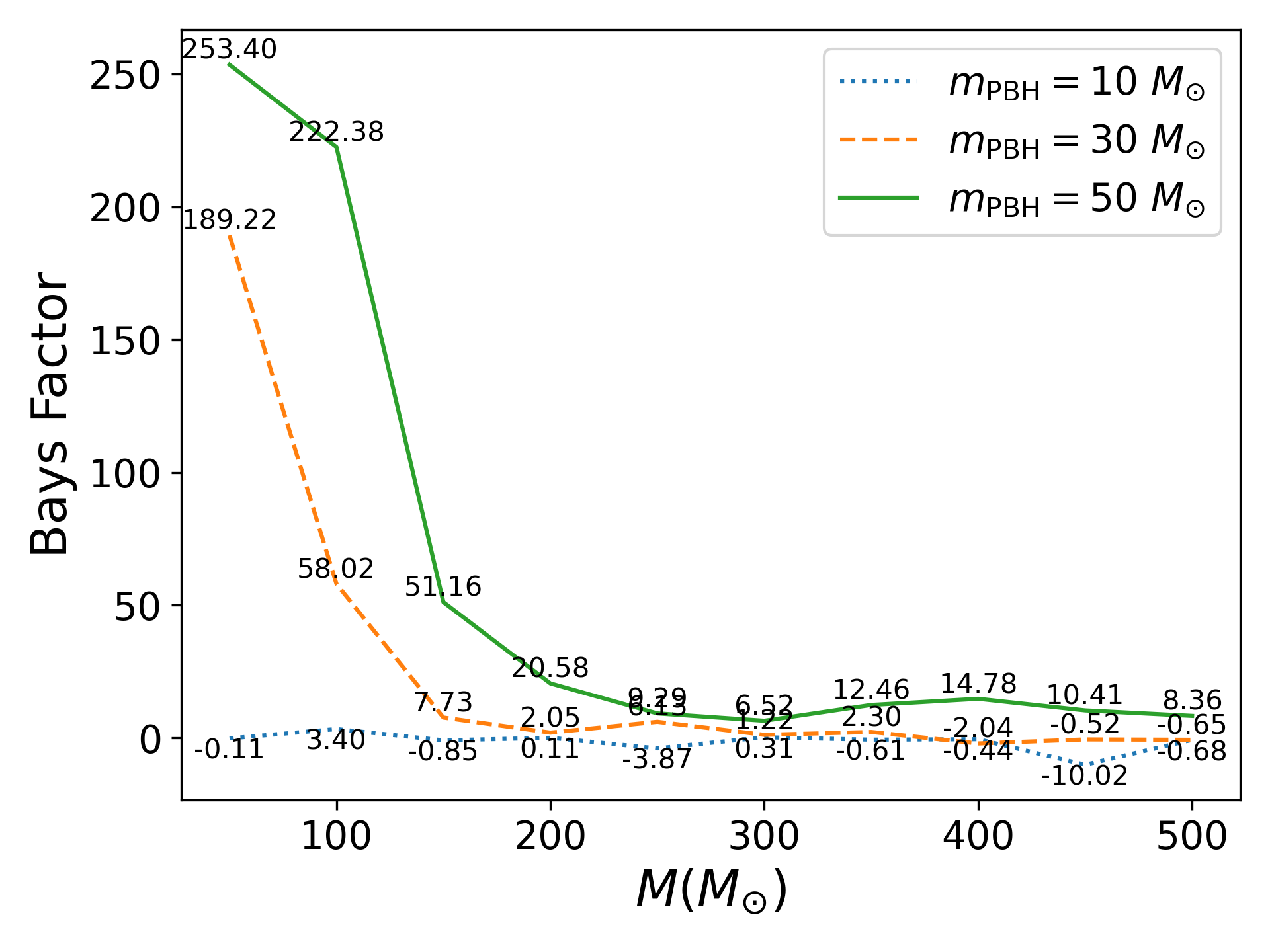}
\caption{The dependence of Bayes factors on $M$. The dotted blue, dashed orange, and solid green lines correspond to the Bayes factors for datasets with $m_{\text{PBH}} = 10~M_{\odot}$, $m_{\text{PBH}} = 30~M_{\odot}$, and $m_{\text{PBH}} = 50~M_{\odot}$, respectively.}
\label{BF_dPBH_m}
\end{figure}

In this section, we consider more general scenarios by computing the Bayes factors while varying the source parameters: the symmetric mass ratio $\eta$ and the total mass $M$. We first vary the total mass $M$, while still assuming that the data is affected by a dPBH. The Bayes factors are shown in Fig. \ref{BF_dPBH_m}. 

For this analysis, we inject data with $\eta = 0.25$, keeping the observation duration, phase of coalescence, redshift, and angular parameters unchanged. The total mass $M$ is varied within the range of $50 \sim 500~M_{\odot}$, with a Bayes factor computed at every $5~M_{\odot}$ increment. The corresponding Bayes factor values are marked in Fig. \ref{BF_dPBH_m}.

Clearly, when $M$ is small, the Bayes factors for $m_{\text{PBH}} = 30~M_{\odot}$ and $m_{\text{PBH}} = 50~M_{\odot}$ are significantly greater than $8$, indicating that the dPBH and PBH detection hypotheses can be distinguished. A smaller total mass $M$ corresponds to a wider frequency range, making it easier to differentiate between these two lens models. Conversely, as $M$ increases, the frequency range narrows, resulting in smaller Bayes factors for larger $M$. 

Additionally, by comparing the three curves, we observe that a larger $m_{\text{PBH}}$ leads to a higher Bayes factor. When $M$ is small and $m_{\text{PBH}}$ is large, the Bayes factor exceeds 200. However, as $M$ increases, even the curve corresponding to a large $m_{\text{PBH}}$ drops rapidly. This occurs because the frequency range decreases sharply with increasing $M$. 

For $M > 200~M_{\odot}$, the Bayes factor remains significantly above $8$ when $m_{\text{PBH}} = 50~M_{\odot}$. However, when $m_{\text{PBH}}$ is smaller, such as $m_{\text{PBH}} = 30~M_{\odot}$ or $m_{\text{PBH}} = 10~M_{\odot}$, the Bayes factor may fall below $8$. In such cases, we are unable to determine which lens detection hypothesis the data favors.

\begin{figure}[h]
\centering
\includegraphics[width=0.5\textwidth]{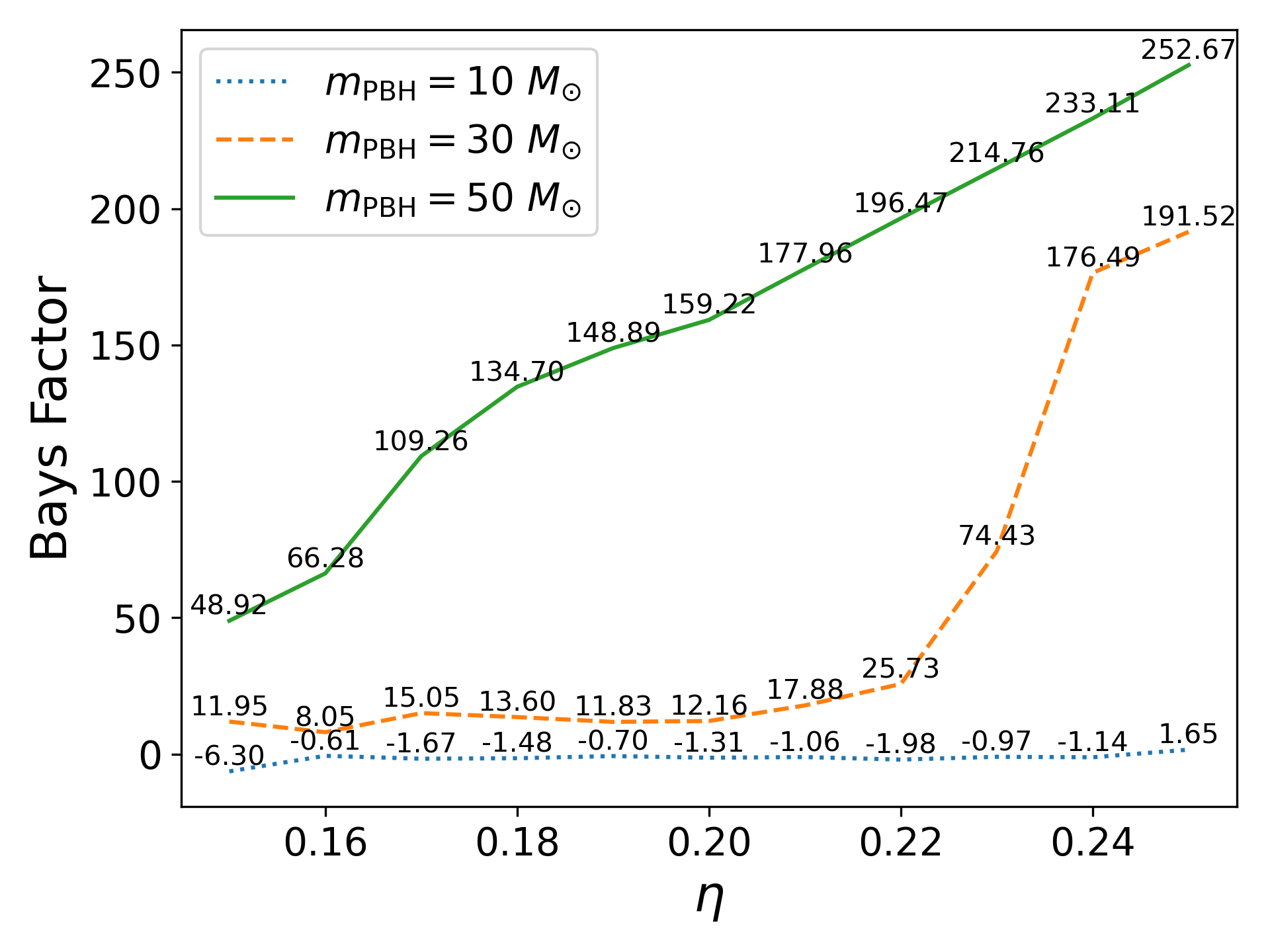}
\caption{The dependence of Bayes factors on $\eta$. The dotted blue, dashed orange, and solid green lines correspond to the Bayes factors for datasets with $m_{\text{PBH}} = 10~M_{\odot}$, $m_{\text{PBH}} = 30~M_{\odot}$, and $m_{\text{PBH}} = 50~M_{\odot}$, respectively.}
\label{BF_dPBH_eta}
\end{figure}

Next, we examine the variation of the Bayes factor with respect to changes in the symmetric mass ratio $\eta$. The range of $\eta$ is set to $0.15 \sim 0.25$. When $\eta = 0.25$, the two black holes in the source system have equal masses. The step size for $\eta$ is chosen as $0.1$. The results are presented in Fig. \ref{BF_dPBH_eta}. 

For this analysis, we inject data with $M = 50~M_{\odot}$ to ensure a wide frequency range, while keeping all other parameters unchanged from previous settings. The corresponding Bayes factor values for different $\eta$ are marked in Fig. \ref{BF_dPBH_eta}.

It is evident that there is a positive correlation between $\eta$ and the Bayes factor. When $\eta = 0.25$, meaning that the two black holes in the binary system have equal masses, the ability to distinguish between the two lens models is maximized. This occurs because a symmetric mass ratio closer to $0.25$ corresponds to a wider frequency range, which in turn leads to a higher Bayes factor. Additionally, a larger $m_{\text{PBH}}$ also results in a higher Bayes factor. 

We observe that when $m_{\text{PBH}} = 10~M_{\odot}$, all Bayes factors remain below $8$. However, when $m_{\text{PBH}} = 30~M_{\odot}$ or $m_{\text{PBH}} = 50~M_{\odot}$, all Bayes factors exceed $8$, even for $\eta = 0.15$. 

Notably, there is a sudden increase in the Bayes factor between $\eta = 0.22$ and $\eta = 0.24$ when $m_{\text{PBH}} = 30~M_{\odot}$. A similar trend occurs between $\eta = 0.15$ and $\eta = 0.18$ when $m_{\text{PBH}} = 50~M_{\odot}$. This phenomenon arises because, for $m_{\text{PBH}} = 10~M_{\odot}$, there exists a similar signal within the prior of $m_{\text{PBH}}$ under the PBH detection hypothesis. In this case, distinguishing between the two signals is challenging due to the limited frequency range. However, when $m_{\text{PBH}} = 30~M_{\odot}$ and $\eta = 0.22$, the frequency range is sufficiently broad, allowing the Bayesian framework to effectively differentiate between the two hypotheses. 

Similarly, for $m_{\text{PBH}} = 50~M_{\odot}$, the sudden rise in the solid orange curve is attributed to the rapid expansion of the frequency range.




\section{Conclusions}
\label{Conclusions}

In this work, we employ the Bayesian inference technique to distinguish between PBHs and PBHs surrounded by particle dark matter, known as dPBHs. The lensing effect of PBHs is computed analytically, while for dPBHs, we evaluate the lensing effect using the asymptotic expansion method in the wave optics regime and adopt the geometrical optics approximation in the intermediate and high-frequency regimes. By combining the lensing effect with the waveform model characterized by IMRPhenomD, we compute the Bayesian inference algorithm using \texttt{dynesty}, which is based on the nested sampling method.

We assume that the detected GW signal is affected by a dPBH lens and consider the ground-based detectors ET and CE in our calculations. Since we lack additional information about the parameters, we adopt flat prior distributions. The priors for intrinsic parameters are centered around their injected values, while extrinsic parameters are sampled over the entire parameter space. Our primary focus is on the posterior distributions of these parameters and the Bayes factor. We set the threshold at $\log \text{BF} = 8$. When $\log \text{BF} \geqslant 8$, we conclude that the data favors the PBH lens detection hypothesis.

We first present the posterior distribution. From the results, we observe that the parameter estimations for the symmetric mass ratio $\eta$, total mass $M$, source redshift $z_S$, and BH mass $m_{\text{PBH}}$ are closely aligned with the injected values. However, the estimation of the observation duration is less accurate due to the weak signal at early times. Additionally, the estimation results for the coalescence phase and other angular parameters exhibit lower accuracy. Nevertheless, these angular parameters are not critical to our analysis. The logarithm of the Bayes factor is $\log \text{BF}_{\text{PBH}}^{\text{dPBH}} = 74.4319$, which is significantly greater than $8$, indicating that Bayesian analysis can effectively distinguish between these two lens models. 

Next, we examine the variation of the Bayes factor with respect to changes in the source parameters $\eta$ and $M$. We first compute the Bayes factors for different values of $M$, presenting three curves corresponding to different injected values of $m_{\text{PBH}}$. We observe that, for the same injected $m_{\text{PBH}}$, the Bayes factor increases as $M$ decreases. This is because a smaller $M$ results in a broader frequency range, which enhances the ability to distinguish between the two lens models, leading to a larger Bayes factor. Subsequently, we compute the Bayes factors for different values of $\eta$. The results indicate that a larger $\eta$ corresponds to a higher Bayes factor. For the same injected $m_{\text{dPBH}}$, the Bayes factor reaches its maximum when $\eta = 0.25$, as the frequency range is widest when the two black holes in the binary system have equal masses. Additionally, we find that, for a fixed $\eta$, the Bayes factor increases with larger $m_{\text{PBH}}$. 

In summary, the ability of the Bayesian framework to distinguish between dPBHs and PBHs improves with a broader frequency range or a larger $m_{\text{PBH}}$.




\section{Acknowledgement}

We are grateful to Han Gil Choi, Liang Dai, Shuo Sun, and Xiangyu Lyu for their helpful discussion. This work is supported by the National Key Research and Development Program of China Grant Nos. 2023YFC2206702, 2023YFC2206703 and 2021YFC2203001; National Natural Science Foundation of China under Grants Nos. 12322301, 12275021, 12073005 and 12021003; the Guangdong Basic and Applied Basic Research Foundation(Grant No. 2023A1515030116); and the Interdiscipline Research Funds of Beijing Normal University.
\bibliography{dPBH}

\begin{thebibliography}{96}%
\makeatletter
\providecommand \@ifxundefined [1]{%
 \@ifx{#1\undefined}
}%
\providecommand \@ifnum [1]{%
 \ifnum #1\expandafter \@firstoftwo
 \else \expandafter \@secondoftwo
 \fi
}%
\providecommand \@ifx [1]{%
 \ifx #1\expandafter \@firstoftwo
 \else \expandafter \@secondoftwo
 \fi
}%
\providecommand \natexlab [1]{#1}%
\providecommand \enquote  [1]{``#1''}%
\providecommand \bibnamefont  [1]{#1}%
\providecommand \bibfnamefont [1]{#1}%
\providecommand \citenamefont [1]{#1}%
\providecommand \href@noop [0]{\@secondoftwo}%
\providecommand \href [0]{\begingroup \@sanitize@url \@href}%
\providecommand \@href[1]{\@@startlink{#1}\@@href}%
\providecommand \@@href[1]{\endgroup#1\@@endlink}%
\providecommand \@sanitize@url [0]{\catcode `\\12\catcode `\$12\catcode
  `\&12\catcode `\#12\catcode `\^12\catcode `\_12\catcode `\%12\relax}%
\providecommand \@@startlink[1]{}%
\providecommand \@@endlink[0]{}%
\providecommand \url  [0]{\begingroup\@sanitize@url \@url }%
\providecommand \@url [1]{\endgroup\@href {#1}{\urlprefix }}%
\providecommand \urlprefix  [0]{URL }%
\providecommand \Eprint [0]{\href }%
\providecommand \doibase [0]{http://dx.doi.org/}%
\providecommand \selectlanguage [0]{\@gobble}%
\providecommand \bibinfo  [0]{\@secondoftwo}%
\providecommand \bibfield  [0]{\@secondoftwo}%
\providecommand \translation [1]{[#1]}%
\providecommand \BibitemOpen [0]{}%
\providecommand \bibitemStop [0]{}%
\providecommand \bibitemNoStop [0]{.\EOS\space}%
\providecommand \EOS [0]{\spacefactor3000\relax}%
\providecommand \BibitemShut  [1]{\csname bibitem#1\endcsname}%
\let\auto@bib@innerbib\@empty
\bibitem [{\citenamefont {Aasi}\ \emph {et~al.}(2015)\citenamefont {Aasi} \emph
  {et~al.}}]{LIGOScientific:2014pky}%
  \BibitemOpen
  \bibfield  {author} {\bibinfo {author} {\bibfnamefont {J.}~\bibnamefont
  {Aasi}} \emph {et~al.} (\bibinfo {collaboration} {LIGO Scientific}),\ }\href
  {\doibase 10.1088/0264-9381/32/7/074001} {\bibfield  {journal} {\bibinfo
  {journal} {Class. Quant. Grav.}\ }\textbf {\bibinfo {volume} {32}},\ \bibinfo
  {pages} {074001} (\bibinfo {year} {2015})},\ \Eprint
  {http://arxiv.org/abs/1411.4547} {arXiv:1411.4547 [gr-qc]} \BibitemShut
  {NoStop}%
\bibitem [{\citenamefont {Abbott}\ \emph
  {et~al.}(2016{\natexlab{a}})\citenamefont {Abbott} \emph
  {et~al.}}]{LIGOScientific:2016emj}%
  \BibitemOpen
  \bibfield  {author} {\bibinfo {author} {\bibfnamefont {B.~P.}\ \bibnamefont
  {Abbott}} \emph {et~al.} (\bibinfo {collaboration} {LIGO Scientific,
  Virgo}),\ }\href {\doibase 10.1103/PhysRevLett.116.131103} {\bibfield
  {journal} {\bibinfo  {journal} {Phys. Rev. Lett.}\ }\textbf {\bibinfo
  {volume} {116}},\ \bibinfo {pages} {131103} (\bibinfo {year}
  {2016}{\natexlab{a}})},\ \Eprint {http://arxiv.org/abs/1602.03838}
  {arXiv:1602.03838 [gr-qc]} \BibitemShut {NoStop}%
\bibitem [{\citenamefont {Acernese}\ \emph {et~al.}(2015)\citenamefont
  {Acernese} \emph {et~al.}}]{VIRGO:2014yos}%
  \BibitemOpen
  \bibfield  {author} {\bibinfo {author} {\bibfnamefont {F.}~\bibnamefont
  {Acernese}} \emph {et~al.} (\bibinfo {collaboration} {VIRGO}),\ }\href
  {\doibase 10.1088/0264-9381/32/2/024001} {\bibfield  {journal} {\bibinfo
  {journal} {Class. Quant. Grav.}\ }\textbf {\bibinfo {volume} {32}},\ \bibinfo
  {pages} {024001} (\bibinfo {year} {2015})},\ \Eprint
  {http://arxiv.org/abs/1408.3978} {arXiv:1408.3978 [gr-qc]} \BibitemShut
  {NoStop}%
\bibitem [{\citenamefont {Abbott}\ \emph
  {et~al.}(2016{\natexlab{b}})\citenamefont {Abbott}, \citenamefont {Abbott},
  \citenamefont {Abbott}, \citenamefont {Abernathy}, \citenamefont {Acernese},
  \citenamefont {Ackley}, \citenamefont {Adams}, \citenamefont {Adams},
  \citenamefont {Addesso},\ and\ \citenamefont {et~al}}]{Abbott:2016ob}%
  \BibitemOpen
  \bibfield  {author} {\bibinfo {author} {\bibfnamefont {B.}~\bibnamefont
  {Abbott}}, \bibinfo {author} {\bibfnamefont {R.}~\bibnamefont {Abbott}},
  \bibinfo {author} {\bibfnamefont {T.}~\bibnamefont {Abbott}}, \bibinfo
  {author} {\bibfnamefont {M.}~\bibnamefont {Abernathy}}, \bibinfo {author}
  {\bibfnamefont {F.}~\bibnamefont {Acernese}}, \bibinfo {author}
  {\bibfnamefont {K.}~\bibnamefont {Ackley}}, \bibinfo {author} {\bibfnamefont
  {C.}~\bibnamefont {Adams}}, \bibinfo {author} {\bibfnamefont
  {T.}~\bibnamefont {Adams}}, \bibinfo {author} {\bibfnamefont
  {P.}~\bibnamefont {Addesso}}, \ and\ \bibinfo {author} {\bibnamefont {et~al}}
  (\bibinfo {collaboration} {LIGO Scientific and VIRGO}),\ }\href {\doibase
  10.1103/PhysRevLett.116.061102} {\bibfield  {journal} {\bibinfo  {journal}
  {Phys. Rev. Lett.}\ }\textbf {\bibinfo {volume} {116}},\ \bibinfo {pages}
  {061102} (\bibinfo {year} {2016}{\natexlab{b}})}\BibitemShut {NoStop}%
\bibitem [{\citenamefont {Narayan}\ \emph {et~al.}(1991)\citenamefont
  {Narayan}, \citenamefont {Piran},\ and\ \citenamefont
  {Shemi}}]{Narayan:1991fn}%
  \BibitemOpen
  \bibfield  {author} {\bibinfo {author} {\bibfnamefont {R.}~\bibnamefont
  {Narayan}}, \bibinfo {author} {\bibfnamefont {T.}~\bibnamefont {Piran}}, \
  and\ \bibinfo {author} {\bibfnamefont {A.}~\bibnamefont {Shemi}},\ }\href
  {\doibase 10.1086/186143} {\bibfield  {journal} {\bibinfo  {journal}
  {Astrophys. J. Lett.}\ }\textbf {\bibinfo {volume} {379}},\ \bibinfo {pages}
  {L17} (\bibinfo {year} {1991})}\BibitemShut {NoStop}%
\bibitem [{\citenamefont {Blanchet}\ \emph {et~al.}(1995)\citenamefont
  {Blanchet}, \citenamefont {Damour}, \citenamefont {Iyer}, \citenamefont
  {Will},\ and\ \citenamefont {Wiseman}}]{Blanchet:1995ez}%
  \BibitemOpen
  \bibfield  {author} {\bibinfo {author} {\bibfnamefont {L.}~\bibnamefont
  {Blanchet}}, \bibinfo {author} {\bibfnamefont {T.}~\bibnamefont {Damour}},
  \bibinfo {author} {\bibfnamefont {B.~R.}\ \bibnamefont {Iyer}}, \bibinfo
  {author} {\bibfnamefont {C.~M.}\ \bibnamefont {Will}}, \ and\ \bibinfo
  {author} {\bibfnamefont {A.~G.}\ \bibnamefont {Wiseman}},\ }\href {\doibase
  10.1103/PhysRevLett.74.3515} {\bibfield  {journal} {\bibinfo  {journal}
  {Phys. Rev. Lett.}\ }\textbf {\bibinfo {volume} {74}},\ \bibinfo {pages}
  {3515} (\bibinfo {year} {1995})},\ \Eprint
  {http://arxiv.org/abs/gr-qc/9501027} {arXiv:gr-qc/9501027} \BibitemShut
  {NoStop}%
\bibitem [{\citenamefont {Akutsu}\ \emph {et~al.}(2021)\citenamefont {Akutsu}
  \emph {et~al.}}]{KAGRA:2020tym}%
  \BibitemOpen
  \bibfield  {author} {\bibinfo {author} {\bibfnamefont {T.}~\bibnamefont
  {Akutsu}} \emph {et~al.} (\bibinfo {collaboration} {KAGRA}),\ }\href
  {\doibase 10.1093/ptep/ptaa125} {\bibfield  {journal} {\bibinfo  {journal}
  {PTEP}\ }\textbf {\bibinfo {volume} {2021}},\ \bibinfo {pages} {05A101}
  (\bibinfo {year} {2021})},\ \Eprint {http://arxiv.org/abs/2005.05574}
  {arXiv:2005.05574 [physics.ins-det]} \BibitemShut {NoStop}%
\bibitem [{\citenamefont {Somiya}(2012)}]{Somiya:2011np}%
  \BibitemOpen
  \bibfield  {author} {\bibinfo {author} {\bibfnamefont {K.}~\bibnamefont
  {Somiya}} (\bibinfo {collaboration} {KAGRA}),\ }\href {\doibase
  10.1088/0264-9381/29/12/124007} {\bibfield  {journal} {\bibinfo  {journal}
  {Class. Quant. Grav.}\ }\textbf {\bibinfo {volume} {29}},\ \bibinfo {pages}
  {124007} (\bibinfo {year} {2012})},\ \Eprint {http://arxiv.org/abs/1111.7185}
  {arXiv:1111.7185 [gr-qc]} \BibitemShut {NoStop}%
\bibitem [{\citenamefont {Aso}\ \emph {et~al.}(2013)\citenamefont {Aso},
  \citenamefont {Michimura}, \citenamefont {Somiya}, \citenamefont {Ando},
  \citenamefont {Miyakawa}, \citenamefont {Sekiguchi}, \citenamefont
  {Tatsumi},\ and\ \citenamefont {Yamamoto}}]{Aso:2013eba}%
  \BibitemOpen
  \bibfield  {author} {\bibinfo {author} {\bibfnamefont {Y.}~\bibnamefont
  {Aso}}, \bibinfo {author} {\bibfnamefont {Y.}~\bibnamefont {Michimura}},
  \bibinfo {author} {\bibfnamefont {K.}~\bibnamefont {Somiya}}, \bibinfo
  {author} {\bibfnamefont {M.}~\bibnamefont {Ando}}, \bibinfo {author}
  {\bibfnamefont {O.}~\bibnamefont {Miyakawa}}, \bibinfo {author}
  {\bibfnamefont {T.}~\bibnamefont {Sekiguchi}}, \bibinfo {author}
  {\bibfnamefont {D.}~\bibnamefont {Tatsumi}}, \ and\ \bibinfo {author}
  {\bibfnamefont {H.}~\bibnamefont {Yamamoto}} (\bibinfo {collaboration}
  {KAGRA}),\ }\href {\doibase 10.1103/PhysRevD.88.043007} {\bibfield  {journal}
  {\bibinfo  {journal} {Phys. Rev. D}\ }\textbf {\bibinfo {volume} {88}},\
  \bibinfo {pages} {043007} (\bibinfo {year} {2013})},\ \Eprint
  {http://arxiv.org/abs/1306.6747} {arXiv:1306.6747 [gr-qc]} \BibitemShut
  {NoStop}%
\bibitem [{\citenamefont {Abbott}\ \emph
  {et~al.}(2023{\natexlab{a}})\citenamefont {Abbott} \emph
  {et~al.}}]{KAGRA:2021vkt}%
  \BibitemOpen
  \bibfield  {author} {\bibinfo {author} {\bibfnamefont {R.}~\bibnamefont
  {Abbott}} \emph {et~al.} (\bibinfo {collaboration} {KAGRA, VIRGO, LIGO
  Scientific}),\ }\href {\doibase 10.1103/PhysRevX.13.041039} {\bibfield
  {journal} {\bibinfo  {journal} {Phys. Rev. X}\ }\textbf {\bibinfo {volume}
  {13}},\ \bibinfo {pages} {041039} (\bibinfo {year} {2023}{\natexlab{a}})},\
  \Eprint {http://arxiv.org/abs/2111.03606} {arXiv:2111.03606 [gr-qc]}
  \BibitemShut {NoStop}%
\bibitem [{\citenamefont {Abbott}\ \emph {et~al.}(2019)\citenamefont {Abbott}
  \emph {et~al.}}]{LIGOScientific:2018mvr}%
  \BibitemOpen
  \bibfield  {author} {\bibinfo {author} {\bibfnamefont {B.~P.}\ \bibnamefont
  {Abbott}} \emph {et~al.} (\bibinfo {collaboration} {LIGO Scientific,
  Virgo}),\ }\href {\doibase 10.1103/PhysRevX.9.031040} {\bibfield  {journal}
  {\bibinfo  {journal} {Phys. Rev. X}\ }\textbf {\bibinfo {volume} {9}},\
  \bibinfo {pages} {031040} (\bibinfo {year} {2019})},\ \Eprint
  {http://arxiv.org/abs/1811.12907} {arXiv:1811.12907 [astro-ph.HE]}
  \BibitemShut {NoStop}%
\bibitem [{\citenamefont {Abbott}\ \emph
  {et~al.}(2021{\natexlab{a}})\citenamefont {Abbott} \emph
  {et~al.}}]{LIGOScientific:2020ibl}%
  \BibitemOpen
  \bibfield  {author} {\bibinfo {author} {\bibfnamefont {R.}~\bibnamefont
  {Abbott}} \emph {et~al.} (\bibinfo {collaboration} {LIGO Scientific,
  Virgo}),\ }\href {\doibase 10.1103/PhysRevX.11.021053} {\bibfield  {journal}
  {\bibinfo  {journal} {Phys. Rev. X}\ }\textbf {\bibinfo {volume} {11}},\
  \bibinfo {pages} {021053} (\bibinfo {year} {2021}{\natexlab{a}})},\ \Eprint
  {http://arxiv.org/abs/2010.14527} {arXiv:2010.14527 [gr-qc]} \BibitemShut
  {NoStop}%
\bibitem [{\citenamefont {Abbott}\ \emph
  {et~al.}(2023{\natexlab{b}})\citenamefont {Abbott} \emph
  {et~al.}}]{LIGOScientific:2021aug}%
  \BibitemOpen
  \bibfield  {author} {\bibinfo {author} {\bibfnamefont {R.}~\bibnamefont
  {Abbott}} \emph {et~al.} (\bibinfo {collaboration} {LIGO Scientific, Virgo,
  KAGRA}),\ }\href {\doibase 10.3847/1538-4357/ac74bb} {\bibfield  {journal}
  {\bibinfo  {journal} {Astrophys. J.}\ }\textbf {\bibinfo {volume} {949}},\
  \bibinfo {pages} {76} (\bibinfo {year} {2023}{\natexlab{b}})},\ \Eprint
  {http://arxiv.org/abs/2111.03604} {arXiv:2111.03604 [astro-ph.CO]}
  \BibitemShut {NoStop}%
\bibitem [{\citenamefont {Abbott}\ \emph {et~al.}(2024)\citenamefont {Abbott}
  \emph {et~al.}}]{LIGOScientific:2021usb}%
  \BibitemOpen
  \bibfield  {author} {\bibinfo {author} {\bibfnamefont {R.}~\bibnamefont
  {Abbott}} \emph {et~al.} (\bibinfo {collaboration} {LIGO Scientific,
  VIRGO}),\ }\href {\doibase 10.1103/PhysRevD.109.022001} {\bibfield  {journal}
  {\bibinfo  {journal} {Phys. Rev. D}\ }\textbf {\bibinfo {volume} {109}},\
  \bibinfo {pages} {022001} (\bibinfo {year} {2024})},\ \Eprint
  {http://arxiv.org/abs/2108.01045} {arXiv:2108.01045 [gr-qc]} \BibitemShut
  {NoStop}%
\bibitem [{\citenamefont {Garr\'on}\ and\ \citenamefont
  {Keitel}(2024)}]{Garron:2023gvd}%
  \BibitemOpen
  \bibfield  {author} {\bibinfo {author} {\bibfnamefont {A.}~\bibnamefont
  {Garr\'on}}\ and\ \bibinfo {author} {\bibfnamefont {D.}~\bibnamefont
  {Keitel}},\ }\href {\doibase 10.1088/1361-6382/ad0b9b} {\bibfield  {journal}
  {\bibinfo  {journal} {Class. Quant. Grav.}\ }\textbf {\bibinfo {volume}
  {41}},\ \bibinfo {pages} {015005} (\bibinfo {year} {2024})},\ \Eprint
  {http://arxiv.org/abs/2306.12908} {arXiv:2306.12908 [gr-qc]} \BibitemShut
  {NoStop}%
\bibitem [{\citenamefont {Ohanian}(1974)}]{Ohanian:1974ys}%
  \BibitemOpen
  \bibfield  {author} {\bibinfo {author} {\bibfnamefont {H.~C.}\ \bibnamefont
  {Ohanian}},\ }\href {\doibase 10.1007/BF01810927} {\bibfield  {journal}
  {\bibinfo  {journal} {Int. J. Theor. Phys.}\ }\textbf {\bibinfo {volume}
  {9}},\ \bibinfo {pages} {425} (\bibinfo {year} {1974})}\BibitemShut {NoStop}%
\bibitem [{\citenamefont {Nakamura}(1998)}]{Nakamura:1997sw}%
  \BibitemOpen
  \bibfield  {author} {\bibinfo {author} {\bibfnamefont {T.~T.}\ \bibnamefont
  {Nakamura}},\ }\href {\doibase 10.1103/PhysRevLett.80.1138} {\bibfield
  {journal} {\bibinfo  {journal} {Phys. Rev. Lett.}\ }\textbf {\bibinfo
  {volume} {80}},\ \bibinfo {pages} {1138} (\bibinfo {year}
  {1998})}\BibitemShut {NoStop}%
\bibitem [{\citenamefont {Nakamura}\ and\ \citenamefont
  {Deguchi}(1999)}]{Nakamura:1999uwi}%
  \BibitemOpen
  \bibfield  {author} {\bibinfo {author} {\bibfnamefont {T.~T.}\ \bibnamefont
  {Nakamura}}\ and\ \bibinfo {author} {\bibfnamefont {S.}~\bibnamefont
  {Deguchi}},\ }\href {\doibase 10.1143/ptps.133.137} {\bibfield  {journal}
  {\bibinfo  {journal} {Prog. Theor. Phys. Suppl.}\ }\textbf {\bibinfo {volume}
  {133}},\ \bibinfo {pages} {137} (\bibinfo {year} {1999})}\BibitemShut
  {NoStop}%
\bibitem [{\citenamefont {Schneider}\ \emph {et~al.}(2006)\citenamefont
  {Schneider}, \citenamefont {Kochanek},\ and\ \citenamefont
  {Wambsganss}}]{Schneider:2006}%
  \BibitemOpen
  \bibfield  {author} {\bibinfo {author} {\bibfnamefont {P.}~\bibnamefont
  {Schneider}}, \bibinfo {author} {\bibfnamefont {C.~S.}\ \bibnamefont
  {Kochanek}}, \ and\ \bibinfo {author} {\bibfnamefont {J.}~\bibnamefont
  {Wambsganss}},\ }\href@noop {} {\emph {\bibinfo {title} {Gravitational
  Lensing: Strong, Weak and Micro}}}\ (\bibinfo {year} {2006})\BibitemShut
  {NoStop}%
\bibitem [{\citenamefont {Takahashi}\ and\ \citenamefont
  {Nakamura}(2003)}]{Takahashi:2003ix}%
  \BibitemOpen
  \bibfield  {author} {\bibinfo {author} {\bibfnamefont {R.}~\bibnamefont
  {Takahashi}}\ and\ \bibinfo {author} {\bibfnamefont {T.}~\bibnamefont
  {Nakamura}},\ }\href {\doibase 10.1086/377430} {\bibfield  {journal}
  {\bibinfo  {journal} {Astrophys. J.}\ }\textbf {\bibinfo {volume} {595}},\
  \bibinfo {pages} {1039} (\bibinfo {year} {2003})},\ \Eprint
  {http://arxiv.org/abs/astro-ph/0305055} {arXiv:astro-ph/0305055} \BibitemShut
  {NoStop}%
\bibitem [{\citenamefont {Oguri}(2019)}]{Oguri:2019fix}%
  \BibitemOpen
  \bibfield  {author} {\bibinfo {author} {\bibfnamefont {M.}~\bibnamefont
  {Oguri}},\ }\href {\doibase 10.1088/1361-6633/ab4fc5} {\bibfield  {journal}
  {\bibinfo  {journal} {Rept. Prog. Phys.}\ }\textbf {\bibinfo {volume} {82}},\
  \bibinfo {pages} {126901} (\bibinfo {year} {2019})},\ \Eprint
  {http://arxiv.org/abs/1907.06830} {arXiv:1907.06830 [astro-ph.CO]}
  \BibitemShut {NoStop}%
\bibitem [{\citenamefont {Leung}\ \emph {et~al.}(2025)\citenamefont {Leung},
  \citenamefont {Jow}, \citenamefont {Saha}, \citenamefont {Dai}, \citenamefont
  {Oguri},\ and\ \citenamefont {Koopmans}}]{Leung:2023lmq}%
  \BibitemOpen
  \bibfield  {author} {\bibinfo {author} {\bibfnamefont {C.}~\bibnamefont
  {Leung}}, \bibinfo {author} {\bibfnamefont {D.}~\bibnamefont {Jow}}, \bibinfo
  {author} {\bibfnamefont {P.}~\bibnamefont {Saha}}, \bibinfo {author}
  {\bibfnamefont {L.}~\bibnamefont {Dai}}, \bibinfo {author} {\bibfnamefont
  {M.}~\bibnamefont {Oguri}}, \ and\ \bibinfo {author} {\bibfnamefont
  {L.~V.~E.}\ \bibnamefont {Koopmans}},\ }\href {\doibase
  10.1007/s11214-025-01157-7} {\bibfield  {journal} {\bibinfo  {journal} {Space
  Sci. Rev.}\ }\textbf {\bibinfo {volume} {221}},\ \bibinfo {pages} {29}
  (\bibinfo {year} {2025})},\ \Eprint {http://arxiv.org/abs/2304.01202}
  {arXiv:2304.01202 [astro-ph.HE]} \BibitemShut {NoStop}%
\bibitem [{\citenamefont {Fan}\ \emph {et~al.}(2017)\citenamefont {Fan},
  \citenamefont {Liao}, \citenamefont {Biesiada}, \citenamefont
  {Piorkowska-Kurpas},\ and\ \citenamefont {Zhu}}]{Fan:2016swi}%
  \BibitemOpen
  \bibfield  {author} {\bibinfo {author} {\bibfnamefont {X.-L.}\ \bibnamefont
  {Fan}}, \bibinfo {author} {\bibfnamefont {K.}~\bibnamefont {Liao}}, \bibinfo
  {author} {\bibfnamefont {M.}~\bibnamefont {Biesiada}}, \bibinfo {author}
  {\bibfnamefont {A.}~\bibnamefont {Piorkowska-Kurpas}}, \ and\ \bibinfo
  {author} {\bibfnamefont {Z.-H.}\ \bibnamefont {Zhu}},\ }\href {\doibase
  10.1103/PhysRevLett.118.091102} {\bibfield  {journal} {\bibinfo  {journal}
  {Phys. Rev. Lett.}\ }\textbf {\bibinfo {volume} {118}},\ \bibinfo {pages}
  {091102} (\bibinfo {year} {2017})},\ \Eprint
  {http://arxiv.org/abs/1612.04095} {arXiv:1612.04095 [gr-qc]} \BibitemShut
  {NoStop}%
\bibitem [{\citenamefont {Broadhurst}\ \emph {et~al.}(2019)\citenamefont
  {Broadhurst}, \citenamefont {Diego},\ and\ \citenamefont
  {Smoot}}]{Broadhurst:2019ijv}%
  \BibitemOpen
  \bibfield  {author} {\bibinfo {author} {\bibfnamefont {T.}~\bibnamefont
  {Broadhurst}}, \bibinfo {author} {\bibfnamefont {J.~M.}\ \bibnamefont
  {Diego}}, \ and\ \bibinfo {author} {\bibfnamefont {G.~F.}\ \bibnamefont
  {Smoot}},\ }\href@noop {} {\  (\bibinfo {year} {2019})},\ \Eprint
  {http://arxiv.org/abs/1901.03190} {arXiv:1901.03190 [astro-ph.CO]}
  \BibitemShut {NoStop}%
\bibitem [{\citenamefont {Singer}\ \emph {et~al.}(2019)\citenamefont {Singer},
  \citenamefont {Goldstein},\ and\ \citenamefont {Bloom}}]{Singer:2019vjs}%
  \BibitemOpen
  \bibfield  {author} {\bibinfo {author} {\bibfnamefont {L.~P.}\ \bibnamefont
  {Singer}}, \bibinfo {author} {\bibfnamefont {D.~A.}\ \bibnamefont
  {Goldstein}}, \ and\ \bibinfo {author} {\bibfnamefont {J.~S.}\ \bibnamefont
  {Bloom}},\ }\href@noop {} {\  (\bibinfo {year} {2019})},\ \Eprint
  {http://arxiv.org/abs/1910.03601} {arXiv:1910.03601 [astro-ph.CO]}
  \BibitemShut {NoStop}%
\bibitem [{\citenamefont {McIsaac}\ \emph {et~al.}(2020)\citenamefont
  {McIsaac}, \citenamefont {Keitel}, \citenamefont {Collett}, \citenamefont
  {Harry}, \citenamefont {Mozzon}, \citenamefont {Edy},\ and\ \citenamefont
  {Bacon}}]{McIsaac:2019use}%
  \BibitemOpen
  \bibfield  {author} {\bibinfo {author} {\bibfnamefont {C.}~\bibnamefont
  {McIsaac}}, \bibinfo {author} {\bibfnamefont {D.}~\bibnamefont {Keitel}},
  \bibinfo {author} {\bibfnamefont {T.}~\bibnamefont {Collett}}, \bibinfo
  {author} {\bibfnamefont {I.}~\bibnamefont {Harry}}, \bibinfo {author}
  {\bibfnamefont {S.}~\bibnamefont {Mozzon}}, \bibinfo {author} {\bibfnamefont
  {O.}~\bibnamefont {Edy}}, \ and\ \bibinfo {author} {\bibfnamefont
  {D.}~\bibnamefont {Bacon}},\ }\href {\doibase 10.1103/PhysRevD.102.084031}
  {\bibfield  {journal} {\bibinfo  {journal} {Phys. Rev. D}\ }\textbf {\bibinfo
  {volume} {102}},\ \bibinfo {pages} {084031} (\bibinfo {year} {2020})},\
  \Eprint {http://arxiv.org/abs/1912.05389} {arXiv:1912.05389 [gr-qc]}
  \BibitemShut {NoStop}%
\bibitem [{\citenamefont {Hannuksela}\ \emph {et~al.}(2019)\citenamefont
  {Hannuksela}, \citenamefont {Haris}, \citenamefont {Ng}, \citenamefont
  {Kumar}, \citenamefont {Mehta}, \citenamefont {Keitel}, \citenamefont {Li},\
  and\ \citenamefont {Ajith}}]{Hannuksela:2019kle}%
  \BibitemOpen
  \bibfield  {author} {\bibinfo {author} {\bibfnamefont {O.~A.}\ \bibnamefont
  {Hannuksela}}, \bibinfo {author} {\bibfnamefont {K.}~\bibnamefont {Haris}},
  \bibinfo {author} {\bibfnamefont {K.~K.~Y.}\ \bibnamefont {Ng}}, \bibinfo
  {author} {\bibfnamefont {S.}~\bibnamefont {Kumar}}, \bibinfo {author}
  {\bibfnamefont {A.~K.}\ \bibnamefont {Mehta}}, \bibinfo {author}
  {\bibfnamefont {D.}~\bibnamefont {Keitel}}, \bibinfo {author} {\bibfnamefont
  {T.~G.~F.}\ \bibnamefont {Li}}, \ and\ \bibinfo {author} {\bibfnamefont
  {P.}~\bibnamefont {Ajith}},\ }\href {\doibase 10.3847/2041-8213/ab0c0f}
  {\bibfield  {journal} {\bibinfo  {journal} {Astrophys. J. Lett.}\ }\textbf
  {\bibinfo {volume} {874}},\ \bibinfo {pages} {L2} (\bibinfo {year} {2019})},\
  \Eprint {http://arxiv.org/abs/1901.02674} {arXiv:1901.02674 [gr-qc]}
  \BibitemShut {NoStop}%
\bibitem [{\citenamefont {Liu}\ \emph {et~al.}(2021)\citenamefont {Liu},
  \citenamefont {Magana~Hernandez},\ and\ \citenamefont
  {Creighton}}]{Liu:2020par}%
  \BibitemOpen
  \bibfield  {author} {\bibinfo {author} {\bibfnamefont {X.}~\bibnamefont
  {Liu}}, \bibinfo {author} {\bibfnamefont {I.}~\bibnamefont
  {Magana~Hernandez}}, \ and\ \bibinfo {author} {\bibfnamefont
  {J.}~\bibnamefont {Creighton}},\ }\href {\doibase 10.3847/1538-4357/abd7eb}
  {\bibfield  {journal} {\bibinfo  {journal} {Astrophys. J.}\ }\textbf
  {\bibinfo {volume} {908}},\ \bibinfo {pages} {97} (\bibinfo {year} {2021})},\
  \Eprint {http://arxiv.org/abs/2009.06539} {arXiv:2009.06539 [astro-ph.HE]}
  \BibitemShut {NoStop}%
\bibitem [{\citenamefont {Dai}\ \emph {et~al.}(2020)\citenamefont {Dai},
  \citenamefont {Zackay}, \citenamefont {Venumadhav}, \citenamefont {Roulet},\
  and\ \citenamefont {Zaldarriaga}}]{Dai:2020tpj}%
  \BibitemOpen
  \bibfield  {author} {\bibinfo {author} {\bibfnamefont {L.}~\bibnamefont
  {Dai}}, \bibinfo {author} {\bibfnamefont {B.}~\bibnamefont {Zackay}},
  \bibinfo {author} {\bibfnamefont {T.}~\bibnamefont {Venumadhav}}, \bibinfo
  {author} {\bibfnamefont {J.}~\bibnamefont {Roulet}}, \ and\ \bibinfo {author}
  {\bibfnamefont {M.}~\bibnamefont {Zaldarriaga}},\ }\href@noop {} {\
  (\bibinfo {year} {2020})},\ \Eprint {http://arxiv.org/abs/2007.12709}
  {arXiv:2007.12709 [astro-ph.HE]} \BibitemShut {NoStop}%
\bibitem [{\citenamefont {Abbott}\ \emph
  {et~al.}(2021{\natexlab{b}})\citenamefont {Abbott} \emph
  {et~al.}}]{LIGOScientific:2021izm}%
  \BibitemOpen
  \bibfield  {author} {\bibinfo {author} {\bibfnamefont {R.}~\bibnamefont
  {Abbott}} \emph {et~al.} (\bibinfo {collaboration} {LIGO Scientific,
  VIRGO}),\ }\href {\doibase 10.3847/1538-4357/ac23db} {\bibfield  {journal}
  {\bibinfo  {journal} {Astrophys. J.}\ }\textbf {\bibinfo {volume} {923}},\
  \bibinfo {pages} {14} (\bibinfo {year} {2021}{\natexlab{b}})},\ \Eprint
  {http://arxiv.org/abs/2105.06384} {arXiv:2105.06384 [gr-qc]} \BibitemShut
  {NoStop}%
\bibitem [{\citenamefont {Diego}\ \emph {et~al.}(2021)\citenamefont {Diego},
  \citenamefont {Broadhurst},\ and\ \citenamefont {Smoot}}]{Diego:2021fyd}%
  \BibitemOpen
  \bibfield  {author} {\bibinfo {author} {\bibfnamefont {J.~M.}\ \bibnamefont
  {Diego}}, \bibinfo {author} {\bibfnamefont {T.}~\bibnamefont {Broadhurst}}, \
  and\ \bibinfo {author} {\bibfnamefont {G.}~\bibnamefont {Smoot}},\ }\href
  {\doibase 10.1103/PhysRevD.104.103529} {\bibfield  {journal} {\bibinfo
  {journal} {Phys. Rev. D}\ }\textbf {\bibinfo {volume} {104}},\ \bibinfo
  {pages} {103529} (\bibinfo {year} {2021})},\ \Eprint
  {http://arxiv.org/abs/2106.06545} {arXiv:2106.06545 [gr-qc]} \BibitemShut
  {NoStop}%
\bibitem [{\citenamefont {Baker}\ and\ \citenamefont
  {Trodden}(2017)}]{Baker:2016reh}%
  \BibitemOpen
  \bibfield  {author} {\bibinfo {author} {\bibfnamefont {T.}~\bibnamefont
  {Baker}}\ and\ \bibinfo {author} {\bibfnamefont {M.}~\bibnamefont
  {Trodden}},\ }\href {\doibase 10.1103/PhysRevD.95.063512} {\bibfield
  {journal} {\bibinfo  {journal} {Phys. Rev. D}\ }\textbf {\bibinfo {volume}
  {95}},\ \bibinfo {pages} {063512} (\bibinfo {year} {2017})},\ \Eprint
  {http://arxiv.org/abs/1612.02004} {arXiv:1612.02004 [astro-ph.CO]}
  \BibitemShut {NoStop}%
\bibitem [{\citenamefont {Goyal}\ \emph {et~al.}(2021)\citenamefont {Goyal},
  \citenamefont {Haris}, \citenamefont {Mehta},\ and\ \citenamefont
  {Ajith}}]{Goyal:2020bkm}%
  \BibitemOpen
  \bibfield  {author} {\bibinfo {author} {\bibfnamefont {S.}~\bibnamefont
  {Goyal}}, \bibinfo {author} {\bibfnamefont {K.}~\bibnamefont {Haris}},
  \bibinfo {author} {\bibfnamefont {A.~K.}\ \bibnamefont {Mehta}}, \ and\
  \bibinfo {author} {\bibfnamefont {P.}~\bibnamefont {Ajith}},\ }\href
  {\doibase 10.1103/PhysRevD.103.024038} {\bibfield  {journal} {\bibinfo
  {journal} {Phys. Rev. D}\ }\textbf {\bibinfo {volume} {103}},\ \bibinfo
  {pages} {024038} (\bibinfo {year} {2021})},\ \Eprint
  {http://arxiv.org/abs/2008.07060} {arXiv:2008.07060 [gr-qc]} \BibitemShut
  {NoStop}%
\bibitem [{\citenamefont {Lai}\ \emph {et~al.}(2018)\citenamefont {Lai},
  \citenamefont {Hannuksela}, \citenamefont {Herrera-Mart\'\i{}n},
  \citenamefont {Diego}, \citenamefont {Broadhurst},\ and\ \citenamefont
  {Li}}]{Lai:2018rto}%
  \BibitemOpen
  \bibfield  {author} {\bibinfo {author} {\bibfnamefont {K.-H.}\ \bibnamefont
  {Lai}}, \bibinfo {author} {\bibfnamefont {O.~A.}\ \bibnamefont {Hannuksela}},
  \bibinfo {author} {\bibfnamefont {A.}~\bibnamefont {Herrera-Mart\'\i{}n}},
  \bibinfo {author} {\bibfnamefont {J.~M.}\ \bibnamefont {Diego}}, \bibinfo
  {author} {\bibfnamefont {T.}~\bibnamefont {Broadhurst}}, \ and\ \bibinfo
  {author} {\bibfnamefont {T.~G.~F.}\ \bibnamefont {Li}},\ }\href {\doibase
  10.1103/PhysRevD.98.083005} {\bibfield  {journal} {\bibinfo  {journal} {Phys.
  Rev. D}\ }\textbf {\bibinfo {volume} {98}},\ \bibinfo {pages} {083005}
  (\bibinfo {year} {2018})},\ \Eprint {http://arxiv.org/abs/1801.07840}
  {arXiv:1801.07840 [gr-qc]} \BibitemShut {NoStop}%
\bibitem [{\citenamefont {Diego}(2020)}]{Diego:2019rzc}%
  \BibitemOpen
  \bibfield  {author} {\bibinfo {author} {\bibfnamefont {J.~M.}\ \bibnamefont
  {Diego}},\ }\href {\doibase 10.1103/PhysRevD.101.123512} {\bibfield
  {journal} {\bibinfo  {journal} {Phys. Rev. D}\ }\textbf {\bibinfo {volume}
  {101}},\ \bibinfo {pages} {123512} (\bibinfo {year} {2020})},\ \Eprint
  {http://arxiv.org/abs/1911.05736} {arXiv:1911.05736 [astro-ph.CO]}
  \BibitemShut {NoStop}%
\bibitem [{\citenamefont {Oguri}\ and\ \citenamefont
  {Takahashi}(2020)}]{Oguri:2020ldf}%
  \BibitemOpen
  \bibfield  {author} {\bibinfo {author} {\bibfnamefont {M.}~\bibnamefont
  {Oguri}}\ and\ \bibinfo {author} {\bibfnamefont {R.}~\bibnamefont
  {Takahashi}},\ }\href {\doibase 10.3847/1538-4357/abafab} {\bibfield
  {journal} {\bibinfo  {journal} {Astrophys. J.}\ }\textbf {\bibinfo {volume}
  {901}},\ \bibinfo {pages} {58} (\bibinfo {year} {2020})},\ \Eprint
  {http://arxiv.org/abs/2007.01936} {arXiv:2007.01936 [astro-ph.CO]}
  \BibitemShut {NoStop}%
\bibitem [{\citenamefont {Xu}\ \emph {et~al.}(2022)\citenamefont {Xu},
  \citenamefont {Ezquiaga},\ and\ \citenamefont {Holz}}]{Xu:2021bfn}%
  \BibitemOpen
  \bibfield  {author} {\bibinfo {author} {\bibfnamefont {F.}~\bibnamefont
  {Xu}}, \bibinfo {author} {\bibfnamefont {J.~M.}\ \bibnamefont {Ezquiaga}}, \
  and\ \bibinfo {author} {\bibfnamefont {D.~E.}\ \bibnamefont {Holz}},\ }\href
  {\doibase 10.3847/1538-4357/ac58f8} {\bibfield  {journal} {\bibinfo
  {journal} {Astrophys. J.}\ }\textbf {\bibinfo {volume} {929}},\ \bibinfo
  {pages} {9} (\bibinfo {year} {2022})},\ \Eprint
  {http://arxiv.org/abs/2105.14390} {arXiv:2105.14390 [astro-ph.CO]}
  \BibitemShut {NoStop}%
\bibitem [{\citenamefont {Tambalo}\ \emph
  {et~al.}(2023{\natexlab{a}})\citenamefont {Tambalo}, \citenamefont
  {Zumalac\'arregui}, \citenamefont {Dai},\ and\ \citenamefont
  {Cheung}}]{Tambalo:2022wlm}%
  \BibitemOpen
  \bibfield  {author} {\bibinfo {author} {\bibfnamefont {G.}~\bibnamefont
  {Tambalo}}, \bibinfo {author} {\bibfnamefont {M.}~\bibnamefont
  {Zumalac\'arregui}}, \bibinfo {author} {\bibfnamefont {L.}~\bibnamefont
  {Dai}}, \ and\ \bibinfo {author} {\bibfnamefont {M.~H.-Y.}\ \bibnamefont
  {Cheung}},\ }\href {\doibase 10.1103/PhysRevD.108.103529} {\bibfield
  {journal} {\bibinfo  {journal} {Phys. Rev. D}\ }\textbf {\bibinfo {volume}
  {108}},\ \bibinfo {pages} {103529} (\bibinfo {year} {2023}{\natexlab{a}})},\
  \Eprint {http://arxiv.org/abs/2212.11960} {arXiv:2212.11960 [astro-ph.CO]}
  \BibitemShut {NoStop}%
\bibitem [{\citenamefont {\c{C}al\i{}\c{s}kan}\ \emph
  {et~al.}(2023)\citenamefont {\c{C}al\i{}\c{s}kan}, \citenamefont {Ji},
  \citenamefont {Cotesta}, \citenamefont {Berti}, \citenamefont
  {Kamionkowski},\ and\ \citenamefont {Marsat}}]{Caliskan:2022hbu}%
  \BibitemOpen
  \bibfield  {author} {\bibinfo {author} {\bibfnamefont {M.}~\bibnamefont
  {\c{C}al\i{}\c{s}kan}}, \bibinfo {author} {\bibfnamefont {L.}~\bibnamefont
  {Ji}}, \bibinfo {author} {\bibfnamefont {R.}~\bibnamefont {Cotesta}},
  \bibinfo {author} {\bibfnamefont {E.}~\bibnamefont {Berti}}, \bibinfo
  {author} {\bibfnamefont {M.}~\bibnamefont {Kamionkowski}}, \ and\ \bibinfo
  {author} {\bibfnamefont {S.}~\bibnamefont {Marsat}},\ }\href {\doibase
  10.1103/PhysRevD.107.043029} {\bibfield  {journal} {\bibinfo  {journal}
  {Phys. Rev. D}\ }\textbf {\bibinfo {volume} {107}},\ \bibinfo {pages}
  {043029} (\bibinfo {year} {2023})},\ \Eprint
  {http://arxiv.org/abs/2206.02803} {arXiv:2206.02803 [astro-ph.CO]}
  \BibitemShut {NoStop}%
\bibitem [{\citenamefont {Urrutia}\ and\ \citenamefont
  {Vaskonen}(2021)}]{Urrutia:2021qak}%
  \BibitemOpen
  \bibfield  {author} {\bibinfo {author} {\bibfnamefont {J.}~\bibnamefont
  {Urrutia}}\ and\ \bibinfo {author} {\bibfnamefont {V.}~\bibnamefont
  {Vaskonen}},\ }\href {\doibase 10.1093/mnras/stab3118} {\bibfield  {journal}
  {\bibinfo  {journal} {Mon. Not. Roy. Astron. Soc.}\ }\textbf {\bibinfo
  {volume} {509}},\ \bibinfo {pages} {1358} (\bibinfo {year} {2021})},\ \Eprint
  {http://arxiv.org/abs/2109.03213} {arXiv:2109.03213 [astro-ph.CO]}
  \BibitemShut {NoStop}%
\bibitem [{\citenamefont {Urrutia}\ \emph {et~al.}(2023)\citenamefont
  {Urrutia}, \citenamefont {Vaskonen},\ and\ \citenamefont
  {Veerm\"ae}}]{Urrutia:2023mtk}%
  \BibitemOpen
  \bibfield  {author} {\bibinfo {author} {\bibfnamefont {J.}~\bibnamefont
  {Urrutia}}, \bibinfo {author} {\bibfnamefont {V.}~\bibnamefont {Vaskonen}}, \
  and\ \bibinfo {author} {\bibfnamefont {H.}~\bibnamefont {Veerm\"ae}},\ }\href
  {\doibase 10.1103/PhysRevD.108.023507} {\bibfield  {journal} {\bibinfo
  {journal} {Phys. Rev. D}\ }\textbf {\bibinfo {volume} {108}},\ \bibinfo
  {pages} {023507} (\bibinfo {year} {2023})},\ \Eprint
  {http://arxiv.org/abs/2303.17601} {arXiv:2303.17601 [astro-ph.CO]}
  \BibitemShut {NoStop}%
\bibitem [{\citenamefont {Frampton}\ \emph {et~al.}(2010)\citenamefont
  {Frampton}, \citenamefont {Kawasaki}, \citenamefont {Takahashi},\ and\
  \citenamefont {Yanagida}}]{Frampton:2010sw}%
  \BibitemOpen
  \bibfield  {author} {\bibinfo {author} {\bibfnamefont {P.~H.}\ \bibnamefont
  {Frampton}}, \bibinfo {author} {\bibfnamefont {M.}~\bibnamefont {Kawasaki}},
  \bibinfo {author} {\bibfnamefont {F.}~\bibnamefont {Takahashi}}, \ and\
  \bibinfo {author} {\bibfnamefont {T.~T.}\ \bibnamefont {Yanagida}},\ }\href
  {\doibase 10.1088/1475-7516/2010/04/023} {\bibfield  {journal} {\bibinfo
  {journal} {JCAP}\ }\textbf {\bibinfo {volume} {04}},\ \bibinfo {pages} {023}
  (\bibinfo {year} {2010})},\ \Eprint {http://arxiv.org/abs/1001.2308}
  {arXiv:1001.2308 [hep-ph]} \BibitemShut {NoStop}%
\bibitem [{\citenamefont {Zhou}\ \emph {et~al.}(2022)\citenamefont {Zhou},
  \citenamefont {Li}, \citenamefont {Liao},\ and\ \citenamefont
  {Huang}}]{Zhou:2022yeo}%
  \BibitemOpen
  \bibfield  {author} {\bibinfo {author} {\bibfnamefont {H.}~\bibnamefont
  {Zhou}}, \bibinfo {author} {\bibfnamefont {Z.}~\bibnamefont {Li}}, \bibinfo
  {author} {\bibfnamefont {K.}~\bibnamefont {Liao}}, \ and\ \bibinfo {author}
  {\bibfnamefont {Z.}~\bibnamefont {Huang}},\ }\href {\doibase
  10.1093/mnras/stac2944} {\bibfield  {journal} {\bibinfo  {journal} {Mon. Not.
  Roy. Astron. Soc.}\ }\textbf {\bibinfo {volume} {518}},\ \bibinfo {pages}
  {149} (\bibinfo {year} {2022})},\ \Eprint {http://arxiv.org/abs/2206.13128}
  {arXiv:2206.13128 [astro-ph.CO]} \BibitemShut {NoStop}%
\bibitem [{\citenamefont {Huang}\ \emph {et~al.}(2023)\citenamefont {Huang},
  \citenamefont {Hu}, \citenamefont {Chen}, \citenamefont {Zhang},
  \citenamefont {Li}, \citenamefont {Gao},\ and\ \citenamefont
  {Lin}}]{Huang:2023prq}%
  \BibitemOpen
  \bibfield  {author} {\bibinfo {author} {\bibfnamefont {S.-J.}\ \bibnamefont
  {Huang}}, \bibinfo {author} {\bibfnamefont {Y.-M.}\ \bibnamefont {Hu}},
  \bibinfo {author} {\bibfnamefont {X.}~\bibnamefont {Chen}}, \bibinfo {author}
  {\bibfnamefont {J.-d.}\ \bibnamefont {Zhang}}, \bibinfo {author}
  {\bibfnamefont {E.-K.}\ \bibnamefont {Li}}, \bibinfo {author} {\bibfnamefont
  {Z.}~\bibnamefont {Gao}}, \ and\ \bibinfo {author} {\bibfnamefont {X.-Y.}\
  \bibnamefont {Lin}},\ }\href {\doibase 10.1088/1475-7516/2023/08/003}
  {\bibfield  {journal} {\bibinfo  {journal} {JCAP}\ }\textbf {\bibinfo
  {volume} {08}},\ \bibinfo {pages} {003} (\bibinfo {year} {2023})},\ \Eprint
  {http://arxiv.org/abs/2304.10435} {arXiv:2304.10435 [astro-ph.CO]}
  \BibitemShut {NoStop}%
\bibitem [{\citenamefont {Huang}\ \emph {et~al.}(2025)\citenamefont {Huang},
  \citenamefont {Li}, \citenamefont {Zhang}, \citenamefont {Chen},
  \citenamefont {Gao}, \citenamefont {Lin},\ and\ \citenamefont
  {Hu}}]{Huang:2024zvk}%
  \BibitemOpen
  \bibfield  {author} {\bibinfo {author} {\bibfnamefont {S.-J.}\ \bibnamefont
  {Huang}}, \bibinfo {author} {\bibfnamefont {E.-K.}\ \bibnamefont {Li}},
  \bibinfo {author} {\bibfnamefont {J.-d.}\ \bibnamefont {Zhang}}, \bibinfo
  {author} {\bibfnamefont {X.}~\bibnamefont {Chen}}, \bibinfo {author}
  {\bibfnamefont {Z.}~\bibnamefont {Gao}}, \bibinfo {author} {\bibfnamefont
  {X.-y.}\ \bibnamefont {Lin}}, \ and\ \bibinfo {author} {\bibfnamefont
  {Y.-M.}\ \bibnamefont {Hu}},\ }\href {\doibase 10.1016/j.dark.2025.101810}
  {\bibfield  {journal} {\bibinfo  {journal} {Phys. Dark Univ.}\ }\textbf
  {\bibinfo {volume} {47}},\ \bibinfo {pages} {101810} (\bibinfo {year}
  {2025})},\ \Eprint {http://arxiv.org/abs/2402.17349} {arXiv:2402.17349
  [astro-ph.CO]} \BibitemShut {NoStop}%
\bibitem [{\citenamefont {Capote}\ \emph {et~al.}(2024)\citenamefont {Capote}
  \emph {et~al.}}]{Capote:2024rmo}%
  \BibitemOpen
  \bibfield  {author} {\bibinfo {author} {\bibfnamefont {E.}~\bibnamefont
  {Capote}} \emph {et~al.},\ }\href@noop {} {\  (\bibinfo {year} {2024})},\
  \Eprint {http://arxiv.org/abs/2411.14607} {arXiv:2411.14607 [gr-qc]}
  \BibitemShut {NoStop}%
\bibitem [{\citenamefont {Branchesi}\ \emph {et~al.}(2023)\citenamefont
  {Branchesi} \emph {et~al.}}]{Branchesi:2023mws}%
  \BibitemOpen
  \bibfield  {author} {\bibinfo {author} {\bibfnamefont {M.}~\bibnamefont
  {Branchesi}} \emph {et~al.},\ }\href {\doibase 10.1088/1475-7516/2023/07/068}
  {\bibfield  {journal} {\bibinfo  {journal} {JCAP}\ }\textbf {\bibinfo
  {volume} {07}},\ \bibinfo {pages} {068} (\bibinfo {year} {2023})},\ \Eprint
  {http://arxiv.org/abs/2303.15923} {arXiv:2303.15923 [gr-qc]} \BibitemShut
  {NoStop}%
\bibitem [{\citenamefont {Punturo}\ \emph {et~al.}(2010)\citenamefont {Punturo}
  \emph {et~al.}}]{Punturo:2010zza}%
  \BibitemOpen
  \bibfield  {author} {\bibinfo {author} {\bibfnamefont {M.}~\bibnamefont
  {Punturo}} \emph {et~al.},\ }\href {\doibase 10.1088/0264-9381/27/8/084007}
  {\bibfield  {journal} {\bibinfo  {journal} {Class. Quant. Grav.}\ }\textbf
  {\bibinfo {volume} {27}},\ \bibinfo {pages} {084007} (\bibinfo {year}
  {2010})}\BibitemShut {NoStop}%
\bibitem [{\citenamefont {Reitze}\ \emph {et~al.}(2019)\citenamefont {Reitze}
  \emph {et~al.}}]{Reitze:2019iox}%
  \BibitemOpen
  \bibfield  {author} {\bibinfo {author} {\bibfnamefont {D.}~\bibnamefont
  {Reitze}} \emph {et~al.},\ }\href@noop {} {\bibfield  {journal} {\bibinfo
  {journal} {Bull. Am. Astron. Soc.}\ }\textbf {\bibinfo {volume} {51}},\
  \bibinfo {pages} {035} (\bibinfo {year} {2019})},\ \Eprint
  {http://arxiv.org/abs/1907.04833} {arXiv:1907.04833 [astro-ph.IM]}
  \BibitemShut {NoStop}%
\bibitem [{\citenamefont {Springel}\ \emph {et~al.}(2005)\citenamefont
  {Springel}, \citenamefont {White}, \citenamefont {Jenkins},\ and\
  \citenamefont {et~al}}]{SpringelV:2005x}%
  \BibitemOpen
  \bibfield  {author} {\bibinfo {author} {\bibfnamefont {V.}~\bibnamefont
  {Springel}}, \bibinfo {author} {\bibfnamefont {S.}~\bibnamefont {White}},
  \bibinfo {author} {\bibfnamefont {A.}~\bibnamefont {Jenkins}}, \ and\
  \bibinfo {author} {\bibnamefont {et~al}},\ }\href {\doibase
  https://doi.org/10.1038/nature03597} {\bibfield  {journal} {\bibinfo
  {journal} {Nature}\ }\textbf {\bibinfo {volume} {435}},\ \bibinfo {pages}
  {629–636} (\bibinfo {year} {2005})}\BibitemShut {NoStop}%
\bibitem [{\citenamefont {Hawking}(1971)}]{Hawking:1971ei}%
  \BibitemOpen
  \bibfield  {author} {\bibinfo {author} {\bibfnamefont {S.}~\bibnamefont
  {Hawking}},\ }\href {\doibase 10.1093/mnras/152.1.75} {\bibfield  {journal}
  {\bibinfo  {journal} {Mon. Not. Roy. Astron. Soc.}\ }\textbf {\bibinfo
  {volume} {152}},\ \bibinfo {pages} {75} (\bibinfo {year} {1971})}\BibitemShut
  {NoStop}%
\bibitem [{\citenamefont {Meszaros}(1975)}]{Meszaros:1975ef}%
  \BibitemOpen
  \bibfield  {author} {\bibinfo {author} {\bibfnamefont {P.}~\bibnamefont
  {Meszaros}},\ }\href@noop {} {\bibfield  {journal} {\bibinfo  {journal}
  {Astron. Astrophys.}\ }\textbf {\bibinfo {volume} {38}},\ \bibinfo {pages}
  {5} (\bibinfo {year} {1975})}\BibitemShut {NoStop}%
\bibitem [{\citenamefont {Garcia-Bellido}\ \emph {et~al.}(1996)\citenamefont
  {Garcia-Bellido}, \citenamefont {Linde},\ and\ \citenamefont
  {Wands}}]{Garcia-Bellido:1996mdl}%
  \BibitemOpen
  \bibfield  {author} {\bibinfo {author} {\bibfnamefont {J.}~\bibnamefont
  {Garcia-Bellido}}, \bibinfo {author} {\bibfnamefont {A.~D.}\ \bibnamefont
  {Linde}}, \ and\ \bibinfo {author} {\bibfnamefont {D.}~\bibnamefont
  {Wands}},\ }\href {\doibase 10.1103/PhysRevD.54.6040} {\bibfield  {journal}
  {\bibinfo  {journal} {Phys. Rev. D}\ }\textbf {\bibinfo {volume} {54}},\
  \bibinfo {pages} {6040} (\bibinfo {year} {1996})},\ \Eprint
  {http://arxiv.org/abs/astro-ph/9605094} {arXiv:astro-ph/9605094} \BibitemShut
  {NoStop}%
\bibitem [{\citenamefont {Boybeyi}\ \emph {et~al.}(2024)\citenamefont
  {Boybeyi}, \citenamefont {Clesse}, \citenamefont {Kuroyanagi},\ and\
  \citenamefont {Sakellariadou}}]{Boybeyi:2024mhp}%
  \BibitemOpen
  \bibfield  {author} {\bibinfo {author} {\bibfnamefont {T.}~\bibnamefont
  {Boybeyi}}, \bibinfo {author} {\bibfnamefont {S.}~\bibnamefont {Clesse}},
  \bibinfo {author} {\bibfnamefont {S.}~\bibnamefont {Kuroyanagi}}, \ and\
  \bibinfo {author} {\bibfnamefont {M.}~\bibnamefont {Sakellariadou}},\
  }\href@noop {} {\  (\bibinfo {year} {2024})},\ \Eprint
  {http://arxiv.org/abs/2412.18318} {arXiv:2412.18318 [astro-ph.CO]}
  \BibitemShut {NoStop}%
\bibitem [{\citenamefont {Carr}\ and\ \citenamefont
  {Hawking}(1974)}]{Carr:1974nx}%
  \BibitemOpen
  \bibfield  {author} {\bibinfo {author} {\bibfnamefont {B.~J.}\ \bibnamefont
  {Carr}}\ and\ \bibinfo {author} {\bibfnamefont {S.~W.}\ \bibnamefont
  {Hawking}},\ }\href {\doibase 10.1093/mnras/168.2.399} {\bibfield  {journal}
  {\bibinfo  {journal} {Mon. Not. Roy. Astron. Soc.}\ }\textbf {\bibinfo
  {volume} {168}},\ \bibinfo {pages} {399} (\bibinfo {year}
  {1974})}\BibitemShut {NoStop}%
\bibitem [{\citenamefont {Carr}(1975)}]{Carr:1975qj}%
  \BibitemOpen
  \bibfield  {author} {\bibinfo {author} {\bibfnamefont {B.~J.}\ \bibnamefont
  {Carr}},\ }\href {\doibase 10.1086/153853} {\bibfield  {journal} {\bibinfo
  {journal} {Astrophys. J.}\ }\textbf {\bibinfo {volume} {201}},\ \bibinfo
  {pages} {1} (\bibinfo {year} {1975})}\BibitemShut {NoStop}%
\bibitem [{\citenamefont {Chapline}(1975)}]{Chapline:1975ojl}%
  \BibitemOpen
  \bibfield  {author} {\bibinfo {author} {\bibfnamefont {G.~F.}\ \bibnamefont
  {Chapline}},\ }\href {\doibase 10.1038/253251a0} {\bibfield  {journal}
  {\bibinfo  {journal} {Nature}\ }\textbf {\bibinfo {volume} {253}},\ \bibinfo
  {pages} {251} (\bibinfo {year} {1975})}\BibitemShut {NoStop}%
\bibitem [{\citenamefont {Kawasaki}\ \emph {et~al.}(1998)\citenamefont
  {Kawasaki}, \citenamefont {Sugiyama},\ and\ \citenamefont
  {Yanagida}}]{Kawasaki:1997ju}%
  \BibitemOpen
  \bibfield  {author} {\bibinfo {author} {\bibfnamefont {M.}~\bibnamefont
  {Kawasaki}}, \bibinfo {author} {\bibfnamefont {N.}~\bibnamefont {Sugiyama}},
  \ and\ \bibinfo {author} {\bibfnamefont {T.}~\bibnamefont {Yanagida}},\
  }\href {\doibase 10.1103/PhysRevD.57.6050} {\bibfield  {journal} {\bibinfo
  {journal} {Phys. Rev. D}\ }\textbf {\bibinfo {volume} {57}},\ \bibinfo
  {pages} {6050} (\bibinfo {year} {1998})},\ \Eprint
  {http://arxiv.org/abs/hep-ph/9710259} {arXiv:hep-ph/9710259} \BibitemShut
  {NoStop}%
\bibitem [{\citenamefont {Kohri}\ \emph {et~al.}(2008)\citenamefont {Kohri},
  \citenamefont {Lyth},\ and\ \citenamefont {Melchiorri}}]{Kohri:2007qn}%
  \BibitemOpen
  \bibfield  {author} {\bibinfo {author} {\bibfnamefont {K.}~\bibnamefont
  {Kohri}}, \bibinfo {author} {\bibfnamefont {D.~H.}\ \bibnamefont {Lyth}}, \
  and\ \bibinfo {author} {\bibfnamefont {A.}~\bibnamefont {Melchiorri}},\
  }\href {\doibase 10.1088/1475-7516/2008/04/038} {\bibfield  {journal}
  {\bibinfo  {journal} {JCAP}\ }\textbf {\bibinfo {volume} {04}},\ \bibinfo
  {pages} {038} (\bibinfo {year} {2008})},\ \Eprint
  {http://arxiv.org/abs/0711.5006} {arXiv:0711.5006 [hep-ph]} \BibitemShut
  {NoStop}%
\bibitem [{\citenamefont {Khlopov}(2010)}]{Khlopov:2008qy}%
  \BibitemOpen
  \bibfield  {author} {\bibinfo {author} {\bibfnamefont {M.~Y.}\ \bibnamefont
  {Khlopov}},\ }\href {\doibase 10.1088/1674-4527/10/6/001} {\bibfield
  {journal} {\bibinfo  {journal} {Res. Astron. Astrophys.}\ }\textbf {\bibinfo
  {volume} {10}},\ \bibinfo {pages} {495} (\bibinfo {year} {2010})},\ \Eprint
  {http://arxiv.org/abs/0801.0116} {arXiv:0801.0116 [astro-ph]} \BibitemShut
  {NoStop}%
\bibitem [{\citenamefont {Carr}\ and\ \citenamefont
  {Kuhnel}(2022)}]{Carr:2021bzv}%
  \BibitemOpen
  \bibfield  {author} {\bibinfo {author} {\bibfnamefont {B.}~\bibnamefont
  {Carr}}\ and\ \bibinfo {author} {\bibfnamefont {F.}~\bibnamefont {Kuhnel}},\
  }\href {\doibase 10.21468/SciPostPhysLectNotes.48} {\bibfield  {journal}
  {\bibinfo  {journal} {SciPost Phys. Lect. Notes}\ }\textbf {\bibinfo {volume}
  {48}},\ \bibinfo {pages} {1} (\bibinfo {year} {2022})},\ \Eprint
  {http://arxiv.org/abs/2110.02821} {arXiv:2110.02821 [astro-ph.CO]}
  \BibitemShut {NoStop}%
\bibitem [{\citenamefont {Green}\ and\ \citenamefont
  {Kavanagh}(2021)}]{Green:2020jor}%
  \BibitemOpen
  \bibfield  {author} {\bibinfo {author} {\bibfnamefont {A.~M.}\ \bibnamefont
  {Green}}\ and\ \bibinfo {author} {\bibfnamefont {B.~J.}\ \bibnamefont
  {Kavanagh}},\ }\href {\doibase 10.1088/1361-6471/abc534} {\bibfield
  {journal} {\bibinfo  {journal} {J. Phys. G}\ }\textbf {\bibinfo {volume}
  {48}},\ \bibinfo {pages} {043001} (\bibinfo {year} {2021})},\ \Eprint
  {http://arxiv.org/abs/2007.10722} {arXiv:2007.10722 [astro-ph.CO]}
  \BibitemShut {NoStop}%
\bibitem [{\citenamefont {Wang}\ \emph
  {et~al.}(2021{\natexlab{a}})\citenamefont {Wang}, \citenamefont
  {Herrera-Mart{\'\i}n},\ and\ \citenamefont {Hu}}]{Wang:2021lij}%
  \BibitemOpen
  \bibfield  {author} {\bibinfo {author} {\bibfnamefont {J.-S.}\ \bibnamefont
  {Wang}}, \bibinfo {author} {\bibfnamefont {A.}~\bibnamefont
  {Herrera-Mart{\'\i}n}}, \ and\ \bibinfo {author} {\bibfnamefont {Y.-M.}\
  \bibnamefont {Hu}},\ }\href {\doibase 10.1103/PhysRevD.104.083515} {\bibfield
   {journal} {\bibinfo  {journal} {Phys. Rev. D}\ }\textbf {\bibinfo {volume}
  {104}},\ \bibinfo {pages} {083515} (\bibinfo {year} {2021}{\natexlab{a}})},\
  \Eprint {http://arxiv.org/abs/2108.12394} {arXiv:2108.12394 [astro-ph.CO]}
  \BibitemShut {NoStop}%
\bibitem [{\citenamefont {Bertschinger}(1985)}]{Bertschinger:1985pd}%
  \BibitemOpen
  \bibfield  {author} {\bibinfo {author} {\bibfnamefont {E.}~\bibnamefont
  {Bertschinger}},\ }\href {\doibase 10.1086/191028} {\bibfield  {journal}
  {\bibinfo  {journal} {Astrophys. J. Suppl.}\ }\textbf {\bibinfo {volume}
  {58}},\ \bibinfo {pages} {39} (\bibinfo {year} {1985})}\BibitemShut {NoStop}%
\bibitem [{\citenamefont {Mack}\ \emph {et~al.}(2007)\citenamefont {Mack},
  \citenamefont {Ostriker},\ and\ \citenamefont {Ricotti}}]{Mack:2006gz}%
  \BibitemOpen
  \bibfield  {author} {\bibinfo {author} {\bibfnamefont {K.~J.}\ \bibnamefont
  {Mack}}, \bibinfo {author} {\bibfnamefont {J.~P.}\ \bibnamefont {Ostriker}},
  \ and\ \bibinfo {author} {\bibfnamefont {M.}~\bibnamefont {Ricotti}},\ }\href
  {\doibase 10.1086/518998} {\bibfield  {journal} {\bibinfo  {journal}
  {Astrophys. J.}\ }\textbf {\bibinfo {volume} {665}},\ \bibinfo {pages} {1277}
  (\bibinfo {year} {2007})},\ \Eprint {http://arxiv.org/abs/astro-ph/0608642}
  {arXiv:astro-ph/0608642} \BibitemShut {NoStop}%
\bibitem [{\citenamefont {Ricotti}\ \emph {et~al.}(2008)\citenamefont
  {Ricotti}, \citenamefont {Ostriker},\ and\ \citenamefont
  {Mack}}]{Ricotti:2007au}%
  \BibitemOpen
  \bibfield  {author} {\bibinfo {author} {\bibfnamefont {M.}~\bibnamefont
  {Ricotti}}, \bibinfo {author} {\bibfnamefont {J.~P.}\ \bibnamefont
  {Ostriker}}, \ and\ \bibinfo {author} {\bibfnamefont {K.~J.}\ \bibnamefont
  {Mack}},\ }\href {\doibase 10.1086/587831} {\bibfield  {journal} {\bibinfo
  {journal} {Astrophys. J.}\ }\textbf {\bibinfo {volume} {680}},\ \bibinfo
  {pages} {829} (\bibinfo {year} {2008})},\ \Eprint
  {http://arxiv.org/abs/0709.0524} {arXiv:0709.0524 [astro-ph]} \BibitemShut
  {NoStop}%
\bibitem [{\citenamefont {Gil~Choi}\ \emph {et~al.}(2024)\citenamefont
  {Gil~Choi}, \citenamefont {Jung}, \citenamefont {Lu},\ and\ \citenamefont
  {Takhistov}}]{GilChoi:2023ahp}%
  \BibitemOpen
  \bibfield  {author} {\bibinfo {author} {\bibfnamefont {H.}~\bibnamefont
  {Gil~Choi}}, \bibinfo {author} {\bibfnamefont {S.}~\bibnamefont {Jung}},
  \bibinfo {author} {\bibfnamefont {P.}~\bibnamefont {Lu}}, \ and\ \bibinfo
  {author} {\bibfnamefont {V.}~\bibnamefont {Takhistov}},\ }\href {\doibase
  10.1103/PhysRevLett.133.101002} {\bibfield  {journal} {\bibinfo  {journal}
  {Phys. Rev. Lett.}\ }\textbf {\bibinfo {volume} {133}},\ \bibinfo {pages}
  {101002} (\bibinfo {year} {2024})},\ \Eprint
  {http://arxiv.org/abs/2311.17829} {arXiv:2311.17829 [astro-ph.CO]}
  \BibitemShut {NoStop}%
\bibitem [{\citenamefont {Jung}\ and\ \citenamefont
  {Shin}(2019)}]{Jung:2017flg}%
  \BibitemOpen
  \bibfield  {author} {\bibinfo {author} {\bibfnamefont {S.}~\bibnamefont
  {Jung}}\ and\ \bibinfo {author} {\bibfnamefont {C.~S.}\ \bibnamefont
  {Shin}},\ }\href {\doibase 10.1103/PhysRevLett.122.041103} {\bibfield
  {journal} {\bibinfo  {journal} {Phys. Rev. Lett.}\ }\textbf {\bibinfo
  {volume} {122}},\ \bibinfo {pages} {041103} (\bibinfo {year} {2019})},\
  \Eprint {http://arxiv.org/abs/1712.01396} {arXiv:1712.01396 [astro-ph.CO]}
  \BibitemShut {NoStop}%
\bibitem [{\citenamefont {Dai}\ \emph {et~al.}(2018)\citenamefont {Dai},
  \citenamefont {Li}, \citenamefont {Zackay}, \citenamefont {Mao},\ and\
  \citenamefont {Lu}}]{Dai:2018enj}%
  \BibitemOpen
  \bibfield  {author} {\bibinfo {author} {\bibfnamefont {L.}~\bibnamefont
  {Dai}}, \bibinfo {author} {\bibfnamefont {S.-S.}\ \bibnamefont {Li}},
  \bibinfo {author} {\bibfnamefont {B.}~\bibnamefont {Zackay}}, \bibinfo
  {author} {\bibfnamefont {S.}~\bibnamefont {Mao}}, \ and\ \bibinfo {author}
  {\bibfnamefont {Y.}~\bibnamefont {Lu}},\ }\href {\doibase
  10.1103/PhysRevD.98.104029} {\bibfield  {journal} {\bibinfo  {journal} {Phys.
  Rev. D}\ }\textbf {\bibinfo {volume} {98}},\ \bibinfo {pages} {104029}
  (\bibinfo {year} {2018})},\ \Eprint {http://arxiv.org/abs/1810.00003}
  {arXiv:1810.00003 [gr-qc]} \BibitemShut {NoStop}%
\bibitem [{\citenamefont {Deguchi}\ and\ \citenamefont
  {Watson}(1986)}]{Deguchi:1986zz}%
  \BibitemOpen
  \bibfield  {author} {\bibinfo {author} {\bibfnamefont {S.}~\bibnamefont
  {Deguchi}}\ and\ \bibinfo {author} {\bibfnamefont {W.~D.}\ \bibnamefont
  {Watson}},\ }\href {\doibase 10.1103/PhysRevD.34.1708} {\bibfield  {journal}
  {\bibinfo  {journal} {Phys. Rev. D}\ }\textbf {\bibinfo {volume} {34}},\
  \bibinfo {pages} {1708} (\bibinfo {year} {1986})}\BibitemShut {NoStop}%
\bibitem [{\citenamefont {Choi}\ \emph {et~al.}(2021)\citenamefont {Choi},
  \citenamefont {Park},\ and\ \citenamefont {Jung}}]{Choi:2021bkx}%
  \BibitemOpen
  \bibfield  {author} {\bibinfo {author} {\bibfnamefont {H.~G.}\ \bibnamefont
  {Choi}}, \bibinfo {author} {\bibfnamefont {C.}~\bibnamefont {Park}}, \ and\
  \bibinfo {author} {\bibfnamefont {S.}~\bibnamefont {Jung}},\ }\href {\doibase
  10.1103/PhysRevD.104.063001} {\bibfield  {journal} {\bibinfo  {journal}
  {Phys. Rev. D}\ }\textbf {\bibinfo {volume} {104}},\ \bibinfo {pages}
  {063001} (\bibinfo {year} {2021})},\ \Eprint
  {http://arxiv.org/abs/2103.08618} {arXiv:2103.08618 [astro-ph.CO]}
  \BibitemShut {NoStop}%
\bibitem [{\citenamefont {Macquart}(2004)}]{Macquart:2004sh}%
  \BibitemOpen
  \bibfield  {author} {\bibinfo {author} {\bibfnamefont {J.-P.}\ \bibnamefont
  {Macquart}},\ }\href {\doibase 10.1051/0004-6361:20034512} {\bibfield
  {journal} {\bibinfo  {journal} {Astron. Astrophys.}\ }\textbf {\bibinfo
  {volume} {422}},\ \bibinfo {pages} {761} (\bibinfo {year} {2004})},\ \Eprint
  {http://arxiv.org/abs/astro-ph/0402661} {arXiv:astro-ph/0402661} \BibitemShut
  {NoStop}%
\bibitem [{\citenamefont {Guo}\ and\ \citenamefont {Lu}(2020)}]{Guo:2020eqw}%
  \BibitemOpen
  \bibfield  {author} {\bibinfo {author} {\bibfnamefont {X.}~\bibnamefont
  {Guo}}\ and\ \bibinfo {author} {\bibfnamefont {Y.}~\bibnamefont {Lu}},\
  }\href {\doibase 10.1103/PhysRevD.102.124076} {\bibfield  {journal} {\bibinfo
   {journal} {Phys. Rev. D}\ }\textbf {\bibinfo {volume} {102}},\ \bibinfo
  {pages} {124076} (\bibinfo {year} {2020})},\ \Eprint
  {http://arxiv.org/abs/2012.03474} {arXiv:2012.03474 [gr-qc]} \BibitemShut
  {NoStop}%
\bibitem [{\citenamefont {Sun}\ and\ \citenamefont {Fan}(2019)}]{Sun:2019ztn}%
  \BibitemOpen
  \bibfield  {author} {\bibinfo {author} {\bibfnamefont {D.}~\bibnamefont
  {Sun}}\ and\ \bibinfo {author} {\bibfnamefont {X.}~\bibnamefont {Fan}},\
  }\href@noop {} {\  (\bibinfo {year} {2019})},\ \Eprint
  {http://arxiv.org/abs/1911.08268} {arXiv:1911.08268 [gr-qc]} \BibitemShut
  {NoStop}%
\bibitem [{\citenamefont {Dai}\ and\ \citenamefont
  {Venumadhav}(2017)}]{Dai:2017huk}%
  \BibitemOpen
  \bibfield  {author} {\bibinfo {author} {\bibfnamefont {L.}~\bibnamefont
  {Dai}}\ and\ \bibinfo {author} {\bibfnamefont {T.}~\bibnamefont
  {Venumadhav}},\ }\href@noop {} {\  (\bibinfo {year} {2017})},\ \Eprint
  {http://arxiv.org/abs/1702.04724} {arXiv:1702.04724 [gr-qc]} \BibitemShut
  {NoStop}%
\bibitem [{\citenamefont {Cremonese}\ \emph {et~al.}(2021)\citenamefont
  {Cremonese}, \citenamefont {Ezquiaga},\ and\ \citenamefont
  {Salzano}}]{Cremonese:2021puh}%
  \BibitemOpen
  \bibfield  {author} {\bibinfo {author} {\bibfnamefont {P.}~\bibnamefont
  {Cremonese}}, \bibinfo {author} {\bibfnamefont {J.~M.}\ \bibnamefont
  {Ezquiaga}}, \ and\ \bibinfo {author} {\bibfnamefont {V.}~\bibnamefont
  {Salzano}},\ }\href {\doibase 10.1103/PhysRevD.104.023503} {\bibfield
  {journal} {\bibinfo  {journal} {Phys. Rev. D}\ }\textbf {\bibinfo {volume}
  {104}},\ \bibinfo {pages} {023503} (\bibinfo {year} {2021})},\ \Eprint
  {http://arxiv.org/abs/2104.07055} {arXiv:2104.07055 [astro-ph.CO]}
  \BibitemShut {NoStop}%
\bibitem [{\citenamefont {Wang}\ \emph
  {et~al.}(2021{\natexlab{b}})\citenamefont {Wang}, \citenamefont {Lo},
  \citenamefont {Li},\ and\ \citenamefont {Chen}}]{Wang:2021kzt}%
  \BibitemOpen
  \bibfield  {author} {\bibinfo {author} {\bibfnamefont {Y.}~\bibnamefont
  {Wang}}, \bibinfo {author} {\bibfnamefont {R.~K.~L.}\ \bibnamefont {Lo}},
  \bibinfo {author} {\bibfnamefont {A.~K.~Y.}\ \bibnamefont {Li}}, \ and\
  \bibinfo {author} {\bibfnamefont {Y.}~\bibnamefont {Chen}},\ }\href {\doibase
  10.1103/PhysRevD.103.104055} {\bibfield  {journal} {\bibinfo  {journal}
  {Phys. Rev. D}\ }\textbf {\bibinfo {volume} {103}},\ \bibinfo {pages}
  {104055} (\bibinfo {year} {2021}{\natexlab{b}})},\ \Eprint
  {http://arxiv.org/abs/2101.08264} {arXiv:2101.08264 [gr-qc]} \BibitemShut
  {NoStop}%
\bibitem [{\citenamefont {Tambalo}\ \emph
  {et~al.}(2023{\natexlab{b}})\citenamefont {Tambalo}, \citenamefont
  {Zumalac\'arregui}, \citenamefont {Dai},\ and\ \citenamefont
  {Cheung}}]{Tambalo:2022plm}%
  \BibitemOpen
  \bibfield  {author} {\bibinfo {author} {\bibfnamefont {G.}~\bibnamefont
  {Tambalo}}, \bibinfo {author} {\bibfnamefont {M.}~\bibnamefont
  {Zumalac\'arregui}}, \bibinfo {author} {\bibfnamefont {L.}~\bibnamefont
  {Dai}}, \ and\ \bibinfo {author} {\bibfnamefont {M.~H.-Y.}\ \bibnamefont
  {Cheung}},\ }\href {\doibase 10.1103/PhysRevD.108.043527} {\bibfield
  {journal} {\bibinfo  {journal} {Phys. Rev. D}\ }\textbf {\bibinfo {volume}
  {108}},\ \bibinfo {pages} {043527} (\bibinfo {year} {2023}{\natexlab{b}})},\
  \Eprint {http://arxiv.org/abs/2210.05658} {arXiv:2210.05658 [gr-qc]}
  \BibitemShut {NoStop}%
\bibitem [{\citenamefont {Takahashi}(2004)}]{Takahashi:2004mc}%
  \BibitemOpen
  \bibfield  {author} {\bibinfo {author} {\bibfnamefont {R.}~\bibnamefont
  {Takahashi}},\ }\href {\doibase 10.1051/0004-6361:20040212} {\bibfield
  {journal} {\bibinfo  {journal} {Astron. Astrophys.}\ }\textbf {\bibinfo
  {volume} {423}},\ \bibinfo {pages} {787} (\bibinfo {year} {2004})},\ \Eprint
  {http://arxiv.org/abs/astro-ph/0402165} {arXiv:astro-ph/0402165} \BibitemShut
  {NoStop}%
\bibitem [{\citenamefont {Morita}\ and\ \citenamefont
  {Soda}(2019)}]{Morita:2019sau}%
  \BibitemOpen
  \bibfield  {author} {\bibinfo {author} {\bibfnamefont {T.}~\bibnamefont
  {Morita}}\ and\ \bibinfo {author} {\bibfnamefont {J.}~\bibnamefont {Soda}},\
  }\href@noop {} {\  (\bibinfo {year} {2019})},\ \Eprint
  {http://arxiv.org/abs/1911.07435} {arXiv:1911.07435 [gr-qc]} \BibitemShut
  {NoStop}%
\bibitem [{\citenamefont {Adamek}\ \emph {et~al.}(2019)\citenamefont {Adamek},
  \citenamefont {Byrnes}, \citenamefont {Gosenca},\ and\ \citenamefont
  {Hotchkiss}}]{Adamek:2019gns}%
  \BibitemOpen
  \bibfield  {author} {\bibinfo {author} {\bibfnamefont {J.}~\bibnamefont
  {Adamek}}, \bibinfo {author} {\bibfnamefont {C.~T.}\ \bibnamefont {Byrnes}},
  \bibinfo {author} {\bibfnamefont {M.}~\bibnamefont {Gosenca}}, \ and\
  \bibinfo {author} {\bibfnamefont {S.}~\bibnamefont {Hotchkiss}},\ }\href
  {\doibase 10.1103/PhysRevD.100.023506} {\bibfield  {journal} {\bibinfo
  {journal} {Phys. Rev. D}\ }\textbf {\bibinfo {volume} {100}},\ \bibinfo
  {pages} {023506} (\bibinfo {year} {2019})},\ \Eprint
  {http://arxiv.org/abs/1901.08528} {arXiv:1901.08528 [astro-ph.CO]}
  \BibitemShut {NoStop}%
\bibitem [{\citenamefont {Berezinsky}\ \emph {et~al.}(2013)\citenamefont
  {Berezinsky}, \citenamefont {Dokuchaev},\ and\ \citenamefont
  {Eroshenko}}]{Berezinsky:2013fxa}%
  \BibitemOpen
  \bibfield  {author} {\bibinfo {author} {\bibfnamefont {V.~S.}\ \bibnamefont
  {Berezinsky}}, \bibinfo {author} {\bibfnamefont {V.~I.}\ \bibnamefont
  {Dokuchaev}}, \ and\ \bibinfo {author} {\bibfnamefont {Y.~N.}\ \bibnamefont
  {Eroshenko}},\ }\href {\doibase 10.1088/1475-7516/2013/11/059} {\bibfield
  {journal} {\bibinfo  {journal} {JCAP}\ }\textbf {\bibinfo {volume} {11}},\
  \bibinfo {pages} {059} (\bibinfo {year} {2013})},\ \Eprint
  {http://arxiv.org/abs/1308.6742} {arXiv:1308.6742 [astro-ph.CO]} \BibitemShut
  {NoStop}%
\bibitem [{\citenamefont {Boudaud}\ \emph {et~al.}(2021)\citenamefont
  {Boudaud}, \citenamefont {Lacroix}, \citenamefont {Stref}, \citenamefont
  {Lavalle},\ and\ \citenamefont {Salati}}]{Boudaud:2021irr}%
  \BibitemOpen
  \bibfield  {author} {\bibinfo {author} {\bibfnamefont {M.}~\bibnamefont
  {Boudaud}}, \bibinfo {author} {\bibfnamefont {T.}~\bibnamefont {Lacroix}},
  \bibinfo {author} {\bibfnamefont {M.}~\bibnamefont {Stref}}, \bibinfo
  {author} {\bibfnamefont {J.}~\bibnamefont {Lavalle}}, \ and\ \bibinfo
  {author} {\bibfnamefont {P.}~\bibnamefont {Salati}},\ }\href {\doibase
  10.1088/1475-7516/2021/08/053} {\bibfield  {journal} {\bibinfo  {journal}
  {JCAP}\ }\textbf {\bibinfo {volume} {08}},\ \bibinfo {pages} {053} (\bibinfo
  {year} {2021})},\ \Eprint {http://arxiv.org/abs/2106.07480} {arXiv:2106.07480
  [astro-ph.CO]} \BibitemShut {NoStop}%
\bibitem [{\citenamefont {Serpico}\ \emph {et~al.}(2020)\citenamefont
  {Serpico}, \citenamefont {Poulin}, \citenamefont {Inman},\ and\ \citenamefont
  {Kohri}}]{Serpico:2020ehh}%
  \BibitemOpen
  \bibfield  {author} {\bibinfo {author} {\bibfnamefont {P.~D.}\ \bibnamefont
  {Serpico}}, \bibinfo {author} {\bibfnamefont {V.}~\bibnamefont {Poulin}},
  \bibinfo {author} {\bibfnamefont {D.}~\bibnamefont {Inman}}, \ and\ \bibinfo
  {author} {\bibfnamefont {K.}~\bibnamefont {Kohri}},\ }\href {\doibase
  10.1103/PhysRevResearch.2.023204} {\bibfield  {journal} {\bibinfo  {journal}
  {Phys. Rev. Res.}\ }\textbf {\bibinfo {volume} {2}},\ \bibinfo {pages}
  {023204} (\bibinfo {year} {2020})},\ \Eprint
  {http://arxiv.org/abs/2002.10771} {arXiv:2002.10771 [astro-ph.CO]}
  \BibitemShut {NoStop}%
\bibitem [{\citenamefont {Lacki}\ and\ \citenamefont
  {Beacom}(2010)}]{Lacki:2010zf}%
  \BibitemOpen
  \bibfield  {author} {\bibinfo {author} {\bibfnamefont {B.~C.}\ \bibnamefont
  {Lacki}}\ and\ \bibinfo {author} {\bibfnamefont {J.~F.}\ \bibnamefont
  {Beacom}},\ }\href {\doibase 10.1088/2041-8205/720/1/L67} {\bibfield
  {journal} {\bibinfo  {journal} {Astrophys. J. Lett.}\ }\textbf {\bibinfo
  {volume} {720}},\ \bibinfo {pages} {L67} (\bibinfo {year} {2010})},\ \Eprint
  {http://arxiv.org/abs/1003.3466} {arXiv:1003.3466 [astro-ph.CO]} \BibitemShut
  {NoStop}%
\bibitem [{\citenamefont {Oguri}\ \emph {et~al.}(2023)\citenamefont {Oguri},
  \citenamefont {Takhistov},\ and\ \citenamefont {Kohri}}]{Oguri:2022fir}%
  \BibitemOpen
  \bibfield  {author} {\bibinfo {author} {\bibfnamefont {M.}~\bibnamefont
  {Oguri}}, \bibinfo {author} {\bibfnamefont {V.}~\bibnamefont {Takhistov}}, \
  and\ \bibinfo {author} {\bibfnamefont {K.}~\bibnamefont {Kohri}},\ }\href
  {\doibase 10.1016/j.physletb.2023.138276} {\bibfield  {journal} {\bibinfo
  {journal} {Phys. Lett. B}\ }\textbf {\bibinfo {volume} {847}},\ \bibinfo
  {pages} {138276} (\bibinfo {year} {2023})},\ \Eprint
  {http://arxiv.org/abs/2208.05957} {arXiv:2208.05957 [astro-ph.CO]}
  \BibitemShut {NoStop}%
\bibitem [{\citenamefont {Finn}(1992)}]{Finn:1992wt}%
  \BibitemOpen
  \bibfield  {author} {\bibinfo {author} {\bibfnamefont {L.~S.}\ \bibnamefont
  {Finn}},\ }\href {\doibase 10.1103/PhysRevD.46.5236} {\bibfield  {journal}
  {\bibinfo  {journal} {Phys. Rev. D}\ }\textbf {\bibinfo {volume} {46}},\
  \bibinfo {pages} {5236} (\bibinfo {year} {1992})},\ \Eprint
  {http://arxiv.org/abs/gr-qc/9209010} {arXiv:gr-qc/9209010} \BibitemShut
  {NoStop}%
\bibitem [{\citenamefont {Cutler}\ and\ \citenamefont
  {Flanagan}(1994)}]{Cutler:1994ys}%
  \BibitemOpen
  \bibfield  {author} {\bibinfo {author} {\bibfnamefont {C.}~\bibnamefont
  {Cutler}}\ and\ \bibinfo {author} {\bibfnamefont {E.~E.}\ \bibnamefont
  {Flanagan}},\ }\href {\doibase 10.1103/PhysRevD.49.2658} {\bibfield
  {journal} {\bibinfo  {journal} {Phys. Rev. D}\ }\textbf {\bibinfo {volume}
  {49}},\ \bibinfo {pages} {2658} (\bibinfo {year} {1994})},\ \Eprint
  {http://arxiv.org/abs/gr-qc/9402014} {arXiv:gr-qc/9402014} \BibitemShut
  {NoStop}%
\bibitem [{\citenamefont {Collaboration}(2023{\natexlab{a}})}]{ET2023}%
  \BibitemOpen
  \bibfield  {author} {\bibinfo {author} {\bibfnamefont {E.~T.}\ \bibnamefont
  {Collaboration}},\ }\href {https://www.et-gw.eu/} {\enquote {\bibinfo {title}
  {Einstein telescope: A third-generation gravitational wave observatory},}\ }
  (\bibinfo {year} {2023}{\natexlab{a}}),\ \bibinfo {note} {accessed:
  2025-02-09}\BibitemShut {NoStop}%
\bibitem [{\citenamefont
  {Collaboration}(2023{\natexlab{b}})}]{CEsensitivity2023}%
  \BibitemOpen
  \bibfield  {author} {\bibinfo {author} {\bibfnamefont {C.~E.}\ \bibnamefont
  {Collaboration}},\ }\href {https://cosmicexplorer.org/sensitivity.html}
  {\enquote {\bibinfo {title} {Cosmic explorer sensitivity curve},}\ }
  (\bibinfo {year} {2023}{\natexlab{b}}),\ \bibinfo {note} {accessed:
  2025-02-09}\BibitemShut {NoStop}%
\bibitem [{\citenamefont {Thrane}\ and\ \citenamefont
  {Talbot}(2019)}]{Thrane:2018qnx}%
  \BibitemOpen
  \bibfield  {author} {\bibinfo {author} {\bibfnamefont {E.}~\bibnamefont
  {Thrane}}\ and\ \bibinfo {author} {\bibfnamefont {C.}~\bibnamefont
  {Talbot}},\ }\href {\doibase 10.1017/pasa.2019.2} {\bibfield  {journal}
  {\bibinfo  {journal} {Publ. Astron. Soc. Austral.}\ }\textbf {\bibinfo
  {volume} {36}},\ \bibinfo {pages} {e010} (\bibinfo {year} {2019})},\ \bibinfo
  {note} {[Erratum: Publ.Astron.Soc.Austral. 37, e036 (2020)]},\ \Eprint
  {http://arxiv.org/abs/1809.02293} {arXiv:1809.02293 [astro-ph.IM]}
  \BibitemShut {NoStop}%
\bibitem [{\citenamefont {Sun}\ \emph {et~al.}(2024)\citenamefont {Sun},
  \citenamefont {Shi}, \citenamefont {Zhang},\ and\ \citenamefont
  {Mei}}]{Sun:2024nut}%
  \BibitemOpen
  \bibfield  {author} {\bibinfo {author} {\bibfnamefont {S.}~\bibnamefont
  {Sun}}, \bibinfo {author} {\bibfnamefont {C.}~\bibnamefont {Shi}}, \bibinfo
  {author} {\bibfnamefont {J.-d.}\ \bibnamefont {Zhang}}, \ and\ \bibinfo
  {author} {\bibfnamefont {J.}~\bibnamefont {Mei}},\ }\href {\doibase
  10.1103/PhysRevD.110.024050} {\bibfield  {journal} {\bibinfo  {journal}
  {Phys. Rev. D}\ }\textbf {\bibinfo {volume} {110}},\ \bibinfo {pages}
  {024050} (\bibinfo {year} {2024})},\ \Eprint
  {http://arxiv.org/abs/2401.11416} {arXiv:2401.11416 [gr-qc]} \BibitemShut
  {NoStop}%
\bibitem [{\citenamefont {Speagle}(2020)}]{Speagle:2019ivv}%
  \BibitemOpen
  \bibfield  {author} {\bibinfo {author} {\bibfnamefont {J.~S.}\ \bibnamefont
  {Speagle}},\ }\href {\doibase 10.1093/mnras/staa278} {\bibfield  {journal}
  {\bibinfo  {journal} {Mon. Not. Roy. Astron. Soc.}\ }\textbf {\bibinfo
  {volume} {493}},\ \bibinfo {pages} {3132} (\bibinfo {year} {2020})},\ \Eprint
  {http://arxiv.org/abs/1904.02180} {arXiv:1904.02180 [astro-ph.IM]}
  \BibitemShut {NoStop}%
\bibitem [{\citenamefont {Zenodo}(2023)}]{zenodo2023}%
  \BibitemOpen
  \bibfield  {author} {\bibinfo {author} {\bibnamefont {Zenodo}},\ }\href
  {https://zenodo.org/records/12537467} {\enquote {\bibinfo {title} {Zenodo
  record 12537467},}\ } (\bibinfo {year} {2023}),\ \bibinfo {note} {accessed:
  2025-02-21}\BibitemShut {NoStop}%
\bibitem [{\citenamefont {Skilling}(2006)}]{Skilling:2006gxv}%
  \BibitemOpen
  \bibfield  {author} {\bibinfo {author} {\bibfnamefont {J.}~\bibnamefont
  {Skilling}},\ }\href {\doibase 10.1214/06-BA127} {\bibfield  {journal}
  {\bibinfo  {journal} {Bayesian Analysis}\ }\textbf {\bibinfo {volume} {1}},\
  \bibinfo {pages} {833} (\bibinfo {year} {2006})}\BibitemShut {NoStop}%
\bibitem [{\citenamefont {Khan}\ \emph {et~al.}(2016)\citenamefont {Khan},
  \citenamefont {Husa}, \citenamefont {Hannam}, \citenamefont {Ohme},
  \citenamefont {P\"urrer}, \citenamefont {Jim\'enez~Forteza},\ and\
  \citenamefont {Boh\'e}}]{Khan:2015jqa}%
  \BibitemOpen
  \bibfield  {author} {\bibinfo {author} {\bibfnamefont {S.}~\bibnamefont
  {Khan}}, \bibinfo {author} {\bibfnamefont {S.}~\bibnamefont {Husa}}, \bibinfo
  {author} {\bibfnamefont {M.}~\bibnamefont {Hannam}}, \bibinfo {author}
  {\bibfnamefont {F.}~\bibnamefont {Ohme}}, \bibinfo {author} {\bibfnamefont
  {M.}~\bibnamefont {P\"urrer}}, \bibinfo {author} {\bibfnamefont
  {X.}~\bibnamefont {Jim\'enez~Forteza}}, \ and\ \bibinfo {author}
  {\bibfnamefont {A.}~\bibnamefont {Boh\'e}},\ }\href {\doibase
  10.1103/PhysRevD.93.044007} {\bibfield  {journal} {\bibinfo  {journal} {Phys.
  Rev. D}\ }\textbf {\bibinfo {volume} {93}},\ \bibinfo {pages} {044007}
  (\bibinfo {year} {2016})},\ \Eprint {http://arxiv.org/abs/1508.07253}
  {arXiv:1508.07253 [gr-qc]} \BibitemShut {NoStop}%
\end{thebibliography}%
\end{document}